\documentclass[a4paper,11pt]{article}
\pdfoutput=1
\usepackage{geometry}
\usepackage[utf8]{inputenc}
\usepackage{graphicx}
\usepackage{amssymb}
\usepackage{textcomp}
\usepackage{amsmath}
\usepackage{tabularx}
\usepackage{bm}
\usepackage{times}
\usepackage{color,xcolor}
\usepackage{slashed}
\usepackage{multirow}
\usepackage{verbatim}
\usepackage{cancel}
\usepackage{subfigure}
\usepackage{comment}
\usepackage{physics}

\def\linkcolor{cyan!70!black}

\usepackage[
colorlinks=true
,urlcolor=\linkcolor
,anchorcolor=\linkcolor
,citecolor=\linkcolor
,filecolor=\linkcolor
,linkcolor=\linkcolor
,menucolor=\linkcolor
,linktocpage=true
,pdfproducer=medialab
,pdfa=true
]{hyperref}

\usepackage{lipsum} 
\usepackage{setspace}
\usepackage[numbers, sort&compress]{natbib}
\newlength{\bibitemsep}
\newlength{\bibparskip}
\let\oldthebibliography\thebibliography
\renewcommand\thebibliography[1]{%
  \oldthebibliography{#1}%
  \setlength{\parskip}{\bibitemsep}%
  \setlength{\itemsep}{\bibparskip}%
}

\usepackage{orcidlink}

\newcommand{\blue}[1]{\textcolor{blue}{#1}}

\def\beq{\begin{equation}}
\def\eeq{\end{equation}}
\def\bea{\begin{eqnarray}}
\def\eea{\end{eqnarray}}

\def \[{\left[}
\def \]{\right]}
\def \({\left(}
\def \){\right)}

\def \l.{\left.}
\def \r.{\right.}

\begin{document}

\vspace{1cm}

\begin{titlepage}

\vspace*{-1.0truecm}
\begin{flushright}
 \end{flushright}
\vspace{0.8truecm}

\begin{center}
\renewcommand{\baselinestretch}{1.8}\normalsize
\boldmath
{\LARGE\textbf{
Probing ALP Lepton Flavour Violation at $\mu$TRISTAN}}
\unboldmath
\end{center}

\vspace{0.4truecm}

\renewcommand*{\thefootnote}{\fnsymbol{footnote}}

\begin{center}

{Lorenzo Calibbi\,\orcidlink{0000-0002-9322-8076}
\,\footnote{\href{mailto:calibbi@nankai.edu.cn}{calibbi@nankai.edu.cn}}, 
Tong Li\,\orcidlink{0000-0002-6437-8542}
\,\footnote{\href{mailto:litong@nankai.edu.cn}{litong@nankai.edu.cn}},
Lopamudra Mukherjee\,\orcidlink{0000-0001-8765-7563}
\,\footnote{\href{mailto:lopamudra.physics@gmail.com}{lopamudra.physics@gmail.com}},
Yiming Yang\,\orcidlink{0000-0002-7441-7648}
\,\footnote{\href{mailto:yangymphys@outlook.com}{yangymphys@outlook.com}}}
\vspace{0.5truecm}

{\small
{\sl School of Physics, Nankai University, Tianjin 300071, China \vspace{0.15truecm}}

}

\vspace*{5mm}
\end{center}

\renewcommand*{\thefootnote}{\arabic{footnote}}
\setcounter{footnote}{0}

\vspace{0.4cm}
\begin{abstract}
\noindent
Axion-like particles (ALPs) with lepton flavour violating (LFV) interactions are predicted within a wide range of flavoured ALP models. The proposed $\mu$TRISTAN high-energy $e^-\mu^+$ and $\mu^+\mu^+$ collider will provide a good opportunity to explore flavour physics in the charged lepton sector. In this work, based on a model-independent effective Lagrangian describing the ALP leptonic interactions, we investigate the potential of $\mu$TRISTAN to probe ALP LFV couplings. We analyse the testability of selected ALP production channels with potential sensitivity at $\mu$TRISTAN, considering different beams and collision energies, including $e^- \mu^+ \to a \gamma$, $e^- \mu^+ \to e^- \tau^+ a$, $\mu^+ \mu^+ \to \mu^+ \tau^+ a$, and $e^- \mu^+ \to \tau^- \mu^+ a$. The produced ALP $a$ is either long-lived or can promptly decay to flavour violating or conserving charged lepton final states. In particular, combining the above LFV ALP production modes with a suitable LFV decay mode, one can identify signatures that are virtually free of Standard Model background.
We show the resulting sensitivity of $\mu$TRISTAN to LFV ALP couplings and compare it with multiple low-energy leptonic constraints and the future improvements thereof. We find that $\mu$TRISTAN can be generally complementary to searches for low-energy LFV processes and measurements of the leptonic magnetic dipole moments and has the capability to explore unconstrained parameter space for ALP masses in the $\mathcal{O}(1)$ to $\mathcal{O}(100)$~GeV range. In the light ALP regime, however, the parameter space that $\mu$TRISTAN is sensitive to, has been already excluded by low-energy searches for LFV decays.
\end{abstract}

\end{titlepage}

\tableofcontents

 
\section{Introduction}

Axions and axion-like particles~(ALPs) are predicted within a wide class of new physics (NP) models.
Specifically, these fields are pseudo Nambu-Goldstone bosons~(PNGB), that is, light pseudoscalars arising from the spontaneous breaking of a global $\text{U}(1)$ symmetry.
Examples are provided by the Peccei-Quinn symmetry~\cite{Peccei:1977hh,Wilczek:1977pj,Weinberg:1977ma}, the lepton number~\cite{Chikashige:1980ui,Gelmini:1980re,Georgi:1981pg}, or a flavour symmetry~\cite{Davidson:1981zd,Wilczek:1982rv,Reiss:1982sq,Davidson:1983fy,Chang:1987hz,Berezhiani:1990jj,Berezhiani:1990wn,Ema:2016ops,Calibbi:2016hwq}, in the context of well-motivated NP frameworks. These models can address some open questions of the Standard Model~(SM) of particle physics, such as the strong CP problem, the origin of neutrino masses, the hierarchical structure of fermion masses and mixing.

Searching for light ALPs is therefore an effective way to test new fundamental dynamics at energy scales possibly beyond the direct reach of high-energy colliders. This appealing connection\,---\,besides a robust dedicated axion experimental programme and a large number of studies on possible cosmological and astrophysical probes of these light fields, for reviews cf.~\cite{Irastorza:2018dyq,DiLuzio:2020wdo,OHare:2024nmr}\,---\,recently motivated a renewed interest also in the potential of axion/ALP searches at flavour~\cite{Izaguirre:2016dfi,Bjorkeroth:2018dzu,CidVidal:2018blh,Perrevoort:2018ttp,Gavela:2019wzg,Merlo:2019anv,Bauer:2019gfk,Albrecht:2019zul,Cornella:2019uxs,MartinCamalich:2020dfe,Endo:2020mev,Calibbi:2020jvd,Escribano:2020wua,DiLuzio:2020oah,Carmona:2021seb,NA62:2021zjw,Guadagnoli:2021fcj,Bauer:2021mvw,Goudzovski:2022vbt,Jho:2022snj,Altmannshofer:2022ckw,Xing:2022rob,Belle-II:2022heu,Hill:2023dym,Knapen:2023zgi,DiLuzio:2023lmd} and collider experiments~\cite{Brivio:2017ije,Bauer:2017ris,Bauer:2018uxu,Zhang:2021sio,Carmona:2022jid,Bao:2022onq,Han:2022mzp,Yue:2022ash,Liu:2022tqn,Lu:2022zbe,Calibbi:2022izs,Lu:2023ryd,Cheung:2023nzg,Alves:2023sdf,Blasi:2023hvb,Yue:2023mjm,Yue:2024xrc,Li:2024thq,Cheung:2024wve}. 

In this article, we study ALPs in the context of the recently proposed $\mu$TRISTAN project~\cite{Hamada:2022mua}. The idea behind $\mu$TRISTAN is to employ in a high-energy collider the mature methods for cooling and focusing positive muon beams that have been developed by the J-PARC muon $g-2$ experiment~\cite{Abe:2019thb}. The $\mu^+$ beams could  be accelerated up to 1~TeV and made to collide with a 30~GeV $e^-$ beam provided by the TRISTAN ring at KEK, resulting in a $\mu^+e^-$ ``Higgs factory''  with centre of mass (c.m.)~energy $\sqrt{s} \simeq 346$~GeV and an expected integrated luminosity $\mathcal{L}=1~\text{ab}^{-1}$. In addition, the machine could be also operated as a $\mu^+\mu^+$ collider with $\sqrt{s} = 2$~TeV, thus constituting a high-energy leptonic ``discovery machine'' already feasible with present-day technologies.
In the original proposal~\cite{Hamada:2022mua}, it was estimated that the $\mu^+ \mu^+$ mode could attain a luminosity $\mathcal{L}=100~\rm{fb}^{-1}$. 
Both setups can be used to study Higgs phyisics~\cite{Hamada:2022mua}, to perform precision  measurements~\cite{Hamada:2022uyn,Chen:2024tqh}, as well as to probe new physics up to the TeV scale~\cite{Bossi:2020yne,Lu:2020dkx,Das:2022mmh,Dev:2023nha,Fridell:2023gjx,Fukuda:2023yui,Lichtenstein:2023iut,Yang:2023ojm,Das:2024gfg,Ding:2024zaj}.

In the following pages, we explore the capability of the $\mu$TRISTAN proposal to search for ALPs with lepton flavour violating~(LFV) interactions. Lepton flavour violation is a generic prediction of a wide range of ALP models~\cite{Calibbi:2020jvd} and, as we are going to show, it would give rise to distinctive signatures virtually affected by no (irreducible) SM background (BG). Furthermore, while LFV ALP couplings are strongly constrained by low-energy LFV experiments if muons and taus can decay \emph{into} an on-shell ALP~$a$\,---\,both for ALPs decaying inside the detector~\cite{Cornella:2019uxs,Bauer:2021mvw} and in the case of invisible ALPs~\cite{Calibbi:2020jvd}\,---\,the LFV interactions of heavier ALPs, for which the decays $\mu \to e \,a$ and $\tau \to \ell \,a$ are kinematically forbidden, are comparatively poorly constrained.
The aim of this work is to systematically study the role that $\mu$TRISTAN could play in this context by assessing its sensitivity to LFV ALP interactions and comparing it with present and future limits set by low-energy LFV and $g-2$ experiments.

The rest of the paper is organised as follows. 
In Section~\ref{sec:LFV-ALP}, we review the general ALP-lepton interaction Lagrangian and discuss the ALP decay length. In Section~\ref{sec:muTRISTAN}, we list and analyse the possible processes sensitive to LFV ALP interactions in $e^-\mu^+$ and $\mu^+\mu^+$ collisions. By computing the cross sections of these processes, we single out a number of ALP production modes that have the potential to be tested at $\mu$TRISTAN: 
$e^- \mu^+ \to a \gamma$, $e^- \mu^+ \to e^- \tau^+ a$, $\mu^+ \mu^+ \to \mu^+ \tau^+ a$, and $e^- \mu^+ \to \tau^- \mu^+ a$.
In Section~\ref{sec:sensitivity}, we present the results of our simulations of these processes and compute the resulting sensitivity to the LFV ALP couplings that could be achieved at $\mu$TRISTAN. Furthermore, we compare these results with the present and future constraints from the relevant low-energy leptonic processes, in particular LFV $\mu$ and $\tau$ decays. 
We summarise our findings and draw our conclusions in Section~\ref{sec:concl}. Useful formulae and experimental information are collected in the Appendices. 


\section{ALP-lepton interactions}
\label{sec:LFV-ALP}

Instead of adopting a specific ALP model, we will work  within the model-independent framework specified by the following effective (dimension-5) Lagrangian\,---\,see e.g.~\cite{Feng:1997tn,Calibbi:2020jvd}\,---\,that describes the interactions of an ALP $a$ with vector and axial-vector leptonic currents in the basis where the leptons are mass eigenstates:
\begin{align}
    \mathcal{L}_{a\ell\ell} \supset \sum_{i} \frac{\partial_\mu a}{2f_a}\,\bar{\ell_i}C_{\ell_i\ell_i}^A\gamma^\mu\gamma_5\ell_i+\sum_{i\ne j} \frac{\partial_\mu a}{2f_a}\,\bar{\ell_i}\gamma^\mu\left(C_{\ell_i\ell_j}^V+C_{\ell_i\ell_j}^A\gamma_5\right)\ell_j\,, 
    \label{eq:L_ALP1}
\end{align}
where $i,j=1,2,3$ are flavour indices with $\ell_i = (e,\mu,\tau)$, $f_a$ is the ALP decay constant, a cut-off scale related to the breaking scale of the $\text{U}(1)$ symmetry, and the dimensionless coefficients $C_{\ell_i\ell_j}^{A,V}$ form hermitian matrices in flavour space, $C_{\ell_i\ell_j}^{A,V\,\ast}=C_{\ell_j\ell_i}^{A,V}$. 
Within a specific model, the coefficients $C_{\ell_i\ell_j}^{A,V}$ are calculable in terms of the lepton $\text{U}(1)$ charges and the lepton fields' rotations to the mass eigenbasis.
In the following, we do not commit to a specific scenario and just take $C_{\ell_i\ell_j}^{A,V}$ as free parameters. 

Upon integrating by parts and employing the equations of motion of the lepton fields, one can show that the above Lagrangian is equivalent, up to a shift of the model-dependent anomalous ALP coupling to photons, to the following dimension-4 scalar/pseudoscalar interactions:
\begin{align}
\mathcal{L}_{a\ell\ell} \supset
-\mathrm{i}\sum_{i,j} \left[\frac{C_{\ell_i\ell_j}^V(m_{\ell_i}-m_{\ell_j})}{{2}f_a}\,a\bar{\ell_i}\ell_j+\frac{C_{\ell_i\ell_j}^A(m_{\ell_i}+m_{\ell_j})}{{2}f_a}\,a\bar{\ell_i}\gamma_5\ell_j\right]\,.
\label{eq:L_ALP2}
\end{align}
From this expression, we can define dimensionless scalar/pseudoscalar couplings: 
\begin{align}
    g_{ij}^{V}\equiv \frac{C_{\ell_i\ell_j}^{V}(m_{\ell_i}- m_{\ell_j})}{2f_a},
    \quad
    g_{ij}^{A}\equiv \frac{C_{\ell_i\ell_j}^{A}(m_{\ell_j}+ m_{\ell_i})}{2f_a},
    \end{align}
which, due to the mass hierarchy $m_{e}\ll m_{\mu}\ll m_{\tau}$, read to a good approximation:
\begin{align}
    g_{ii}^{A}= \frac{C_{\ell_i\ell_i}^Am_{\ell_i}}{f_a},
    \quad 
    g_{ij}^{A,V}\simeq \pm\frac{C_{\ell_i\ell_j}^{A,V}m_{\ell_j}}{2f_a}~~~(i<j)\,.
    \label{eq:gVA}    
\end{align}

From Eq.~\eqref{eq:L_ALP2}, one can see that the strength of the interactions is approximately proportional to the mass of the heaviest lepton involved. This expression also shows why flavour-conserving vector currents are unphysical and were consequently not introduced in Eq.~\eqref{eq:L_ALP1}.

Throughout the paper, we do not consider ALP fields that are genuine massless Nambu-Goldstone boson. Instead, we introduce a mass term $\frac12 m_a^2 a^2$, assuming that it stems from explicit $\text{U}(1)$ breaking, and we treat the ALP mass $m_a$ as a free parameter. 
Following from the above interactions, it is then clear that, if kinematics allows, the ALP will decay into lepton pairs. The resulting width reads~\cite{Calibbi:2020jvd}:
\begin{align}
%
\Gamma(a\to \bar{\ell_i}\ell_j)=~&\frac{m_a}{8\pi}\sqrt{\left(1-\frac{(m_{\ell_i}+m_{\ell_j})^2}{m_a^2}\right)\left(1-\frac{(m_{\ell_i}-m_{\ell_j})^2}{m_a^2}\right)} \,\times \nonumber \\
& \left[|g^V_{ij}|^2 \left(1-\frac{(m_{\ell_i}+m_{\ell_j})^2}{m_a^2}\right) +
|g^A_{ij}|^2 \left(1-\frac{(m_{\ell_i}-m_{\ell_j})^2}{m_a^2}\right)\right]\,.
\label{eq:atoellell}
\end{align}
For $m_a\gg m_{\ell_i}+m_{\ell_j}$, this decay width corresponds to the following ALP proper decay length:
\begin{align}
c\tau(a\to \ell_i\ell_j) \approx & {16\pi\over m_a } {1\over |g^V_{ij}|^2+|g^A_{ij}|^2}\approx 10^{-6}~{\rm m}~\left({1~{\rm GeV}\over m_a}\right) \left({10^{-8}\over |g^V_{ij}|^2+|g^A_{ij}|^2}\right)\;,
\label{eq:ctau}
\end{align}
where we have summed over the final state charges. As we will see in the following sections, we expect $\mu$TRISTAN to be able to probe LFV ALP couplings as small as $|g^{A,V}_{ij}| \sim 10^{-3} - 10^{-4}$. Hence, Eq.~\eqref{eq:ctau} implies that ALPs whose decays into charged leptons are allowed will always decay promptly inside the detector for the range of couplings we are interested in. Therefore, in this case, ALPs can be searched through distinctive nearly-background-free LFV signals.

\begin{figure}[t]
    \centering
    \subfigure[LFV ($e$-$\mu$) couplings only]{
    \includegraphics[width=0.49\textwidth]{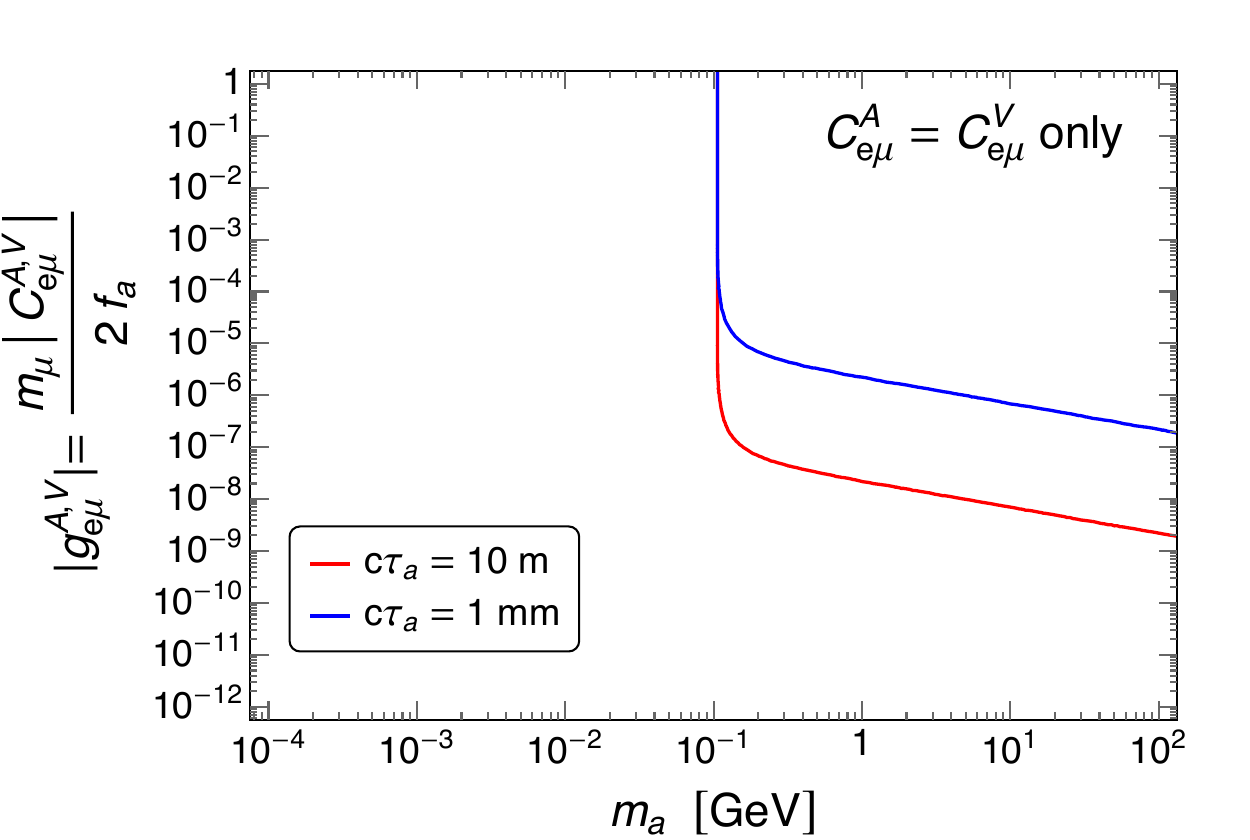}}
    \hfill
    \subfigure[LFV and LFC couplings]{
    \includegraphics[width=0.49\textwidth]{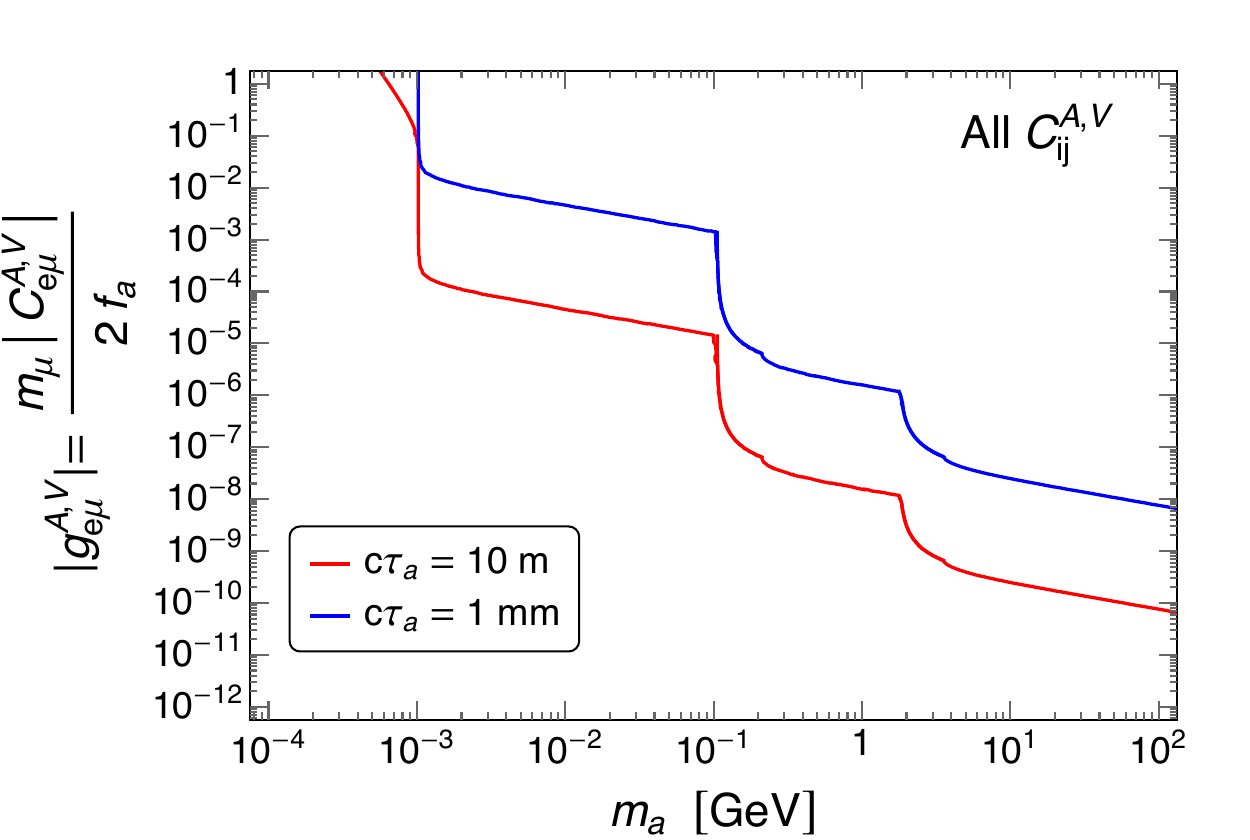}}
    \caption{Contours in the plane $m_a-|g_{ij}^{A,V}|~(= m_{\ell_j}\,|C_{\ell_i\ell_j}^{A,V}|/f_a)$ of the proper decay length $c\tau_a$ of an ALP whose only non-vanishing couplings are $C_{e\mu}^{V}=C_{e\mu}^{A}$ (left panel) 
    and an ALP with all couplings $C_{\ell_i\ell_j}^{A,V}$ ($i\neq j$) and $C_{\ell_i\ell_i}^{A}$ equal (right panel).}
    \label{fig:alplifetime1}
\end{figure}

For $m_a<2\,m_e$ or, if the model does not admit flavour-conserving interactions, $m_a<m_e + m_\mu$,
the ALP can still decay into neutrinos\,---\,either because of a direct coupling or through $W$ exchange at one loop\,---\,but such process is extremely suppressed by small neutrino masses and irrelevant in our context. 
In addition, even if an UV interaction with photons is not present, lepton flavour-conserving (LFC) couplings necessarily induce $a\to\gamma\gamma$.
The decay width for this mode is reported in Appendix~\ref{app:gamma}. %

 We now illustrate the above discussion showing the ALP proper decay length for different scenarios. In the left panel of Figure~\ref{fig:alplifetime1}, the ALP has only LFV interactions to electrons and muons with $C_{e\mu}^{V}=C_{e\mu}^{A}$.
In this case, the ALP is always long-lived below the $a\to e\mu$ kinematic threshold, always short-lived above it unless the coupling $|g^{A,V}_{e\mu}|$ is smaller than $10^{-5}-10^{-6}$, values beyond the $\mu$TRISTAN reach, as we will see. 

In the right plot of Figure~\ref{fig:alplifetime1}, we consider an ALP with all the (LFC and LFV) coefficients $C_{\ell_i \ell_j}^{A,V}$ set to be equal. As we can see, below the $m_\mu + m_e$ threshold, the ALP tends to decay into $e^+e^-$ inside the detector in the large coupling regime, while an ALP lighter than $2\,m_e$  always escapes from the detector appearing as missing energy even if the flavour-conserving couplings inducing $a\to\gamma\gamma$ are present.\footnote{Here we set the model-dependent UV ALP-photon coupling $E_\text{UV}$ to zero, see~Appendix~\ref{app:gamma}.} 
Although this would give an interesting signature at colliders, the $\mu$TRISTAN sensitivity to such light ALPs compares unfavourably with the stringent constraints from searches for LFV $\mu$ and $\tau$ decays, as we will show in the following sections.

\section{LFV processes at $\mu$TRISTAN}
\label{sec:muTRISTAN}

In this section, we examine several processes involving LFV ALPs that could be probed at $\mu$TRISTAN either in the $e^-\mu^+$ or the $\mu^+\mu^+$ mode. 
We first present ALP-mediated LFV processes and then processes with a final-state ALP produced through its LFV interactions. 
After implementing the LFV ALP Lagrangian introduced in the previous section in \textsc{FeynRules}~\cite{Degrande:2011ua,Alloul:2013bka}, we compute the cross sections of the processes we are interested in by means of \textsc{Madgraph}~\cite{Alwall:2011uj,Maltoni:2002qb}. All numbers reported in this section refer to the leading-order cross sections after the standard generator-level cuts on the transverse momentum ($|p_T|>10~\text{GeV}$) and pseudo-rapidity ($|\eta|<2.5$) of all visible final state particles are applied. 

\subsection{ALP-mediated Lepton Flavour Violation}
\label{sec:ALPmediation}

\begin{table}[t]
    \centering
    \renewcommand{\arraystretch}{1.4}
    \begin{tabular}{cccc}
    \hline\hline      
        Process & ALP couplings & Cross section~(fb)  \\
         \hline\hline
        \multirow{2}{*}{$e^- \mu^+ \to e^+ \mu^-$} & $C^A_{e\mu}$~~or~~$C^V_{e\mu}$ & $9.7 \times 10^{-9}$\\
         & $C^A_{e\mu} = C^V_{e\mu}$ & $5.1 \times 10^{-8}$\\
        \hline
        \multirow{2}{*}{$e^- \mu^+ \to \gamma \gamma$} & $C^A_{e\mu}$~~or~~$C^V_{e\mu}$,~~$C_\gamma$  & $4.3 \times 10^{-7}$\\
         & $C^A_{e\mu} = C^V_{e\mu}$,~~$C_\gamma$ & $8.5 \times 10^{-7}$   \\
        \hline
        \multirow{2}{*}{$e^- \mu^+ \to \mu^- \tau^+$} & $C^A_{\mu\tau} = C^A_{e\mu}$~~or~~$C^V_{\mu\tau} = C^V_{e\mu}$ &   $3.1\times 10^{-6}$\\
         & $C^A_{\mu\tau} =C^V_{\mu\tau} = C^A_{e\mu} = C^V_{e\mu}$ & $7.3\times 10^{-6}$ \\
        \hline
        \multirow{2}{*}{$e^- \mu^+ \to \tau^- \tau^+$} & $C^A_{\mu\tau} = C^A_{e\tau}$~~or~~$C^V_{\mu\tau} = C^V_{e\tau}$ & $2.9\times 10^{-4}$ \\
        & $ C^A_{\mu\tau} = C^V_{\mu\tau} = C^A_{e\tau} = C^V_{e\tau}$ & $1.0\times 10^{-3}$ \\
        \hline\hline
                 \multirow{2}{*}{$\mu^+ \mu^+ \to e^+ e^+$} & {$C^A_{e\mu}$~~or~~$C^V_{e\mu}$} &  $2.3 \times 10^{-10}$\\
         & $ C^A_{e\mu} = C^V_{e\mu}$ & $1.2 \times 10^{-9}$\\
        \hline
        \multirow{2}{*}{$\mu^+ \mu^+ \to \tau^+ \tau^+$} & $C^A_{\tau\mu}$~~or~~$C^V_{\tau\mu}$ & $2.3 \times 10^{-5}$\\
         & $C^A_{\tau\mu} = C^V_{\tau\mu}$ & $9.4\times 10^{-5}$\\
        \hline\hline
    \end{tabular}
    \caption{  Cross sections of some ALP-mediated LFV processes in $e^-\mu^+$ collisions (with beam energies $E_e = 30$~GeV, $E_\mu = 1$~TeV) and $\mu^+\mu^+$ collisions (with $E_\mu = 1$~TeV for both beams). For the ALP mass and decay constant we take $m_a = 1$~GeV, $f_a = 100$~GeV. The only non-vanishing couplings are indicated in the second column and are set equal to one, $C^{A,V}_{\ell_i\ell_j}=1$, with the exception of the coupling to photons, $C_\gamma \frac{a}{f_a}F\widetilde F$, that is set to $C_\gamma = \alpha/4\pi$\,---\,that corresponds to $E_\text{UV}=1$ according to the notation in Appendix~\ref{app:gamma}. All cross sections scale as $(100~\text{GeV}/f_a)^4$.
    }
    \label{tab:process_2fa}
\end{table}

In Table~\ref{tab:process_2fa}, we display a number of LFV processes that can be induced in $e^-\mu^+$ and $\mu^+\mu^+$ collisions by an ALP exchange in the $t$ channel or\,---\,only in the case of $e^-\mu^+\to\gamma \gamma$\,---\,in the $s$ channel. 
The couplings are set as indicated in the table and, for the sake of illustration, we take $m_a=1$~GeV. We checked that the $t$-channel processes have a weak dependence on the ALP mass.
These LFV processes could be in principle sensitive to a wide range of ALP masses, even with $m_a$ larger than the c.m.~energy $\sqrt{s}$ at which $\mu$TRISTAN is operated, and are virtually free of any irreducible SM background. However, all processes require two ALP vertices and thus suffer a $\sim 1/f_a^4$ suppression. Hence, as we can see, the resulting cross sections are in general too small to be realistically probed at $\mu$TRISTAN, even assuming that an integrated luminosity as large as $\mathcal{L}=10~\text{ab}^{-1}$ can be achieved. 

Notice that we set $f_a = 100$~GeV and the relevant couplings $C^{A,V}_{\ell_i\ell_j}=1$, which corresponds to $|g^{A,V}_{e\mu}| = 5.3\times 10^{-4}$ and $|g^{A,V}_{\ell\tau}| = 8.9\times 10^{-3}$, 
as one can see from Eq.~\eqref{eq:gVA}.
As we mentioned, the scale $f_a$ is related to the vacuum expectation value (vev) of the scalar field that breaks the underlying $\text{U}(1)$ symmetry. As a consequence, we expect that viable UV completions of the effective Lagrangian in Eq.~\eqref{eq:L_ALP1} will require $f_a\gtrsim 100$~GeV, hence $|g^{A,V}_{e\mu}| \lesssim 10^{-3}$ and $|g^{A,V}_{\ell\tau}| \lesssim 10^{-2}$. 
Hence, the cross sections in Table~\ref{tab:process_2fa} are arguably the largest possible within realistic ALP models and we can conclude that $\mu$TRISTAN would not be capable to test LFV ALPs through these processes.

\subsection{ALP production through LFV interactions}
\label{sec:ALPproduction}

\begin{table}[t!]
    \centering
        \renewcommand{\arraystretch}{1.2}
    \begin{tabular}{cccc}
    \hline\hline
         Process & \quad\quad ALP couplings \quad\quad & \quad $m_a$ (GeV) \quad & \quad Cross section (fb) \quad   \\
    \hline\hline
        \multirow{3}{*}{\blue{$e^- \mu^+ \to a \gamma$}} &
        \multirow{3}{*}{$C^A_{e\mu} = C^V_{e\mu}$} 
         & 1 & $3.0 \times 10^{-2}$\\
         & & 10 & $3.0 \times 10^{-2}$ \\
         & & 100 & $3.3 \times 10^{-2}$ \\
    \hline
    \multirow{3}{*}{$e^- \mu^+ \to a Z$} &
        \multirow{3}{*}{$C^A_{e\mu} = C^V_{e\mu}$} 
         & 1 & $5.9\times 10^{-2}$ \\
         & & 10 & $4.8\times 10^{-2}$ \\
         & & 100 &  $3.1\times 10^{-2}$ \\
    \hline
        \multirow{3}{*}{$e^- \mu^+ \to \mu^- \mu^+ a$} & \multirow{3}{*}{$C^A_{e\mu} = C^V_{e\mu}$} & 1 & $1.4\times 10^{-3}$\\
         & & 10 & $1.1\times 10^{-3}$\\
         & & 100 & $7.2\times 10^{-4}$ \\
    \hline
         \multirow{3}{*}{$e^- \mu^+ \to e^- e^+ a$} & \multirow{3}{*}{$C^A_{e\mu} = C^V_{e\mu}$} & 1 & $8.0\times 10^{-3}$\\
         & & 10 & $4.2\times 10^{-3}$ \\
         & & 100 & $9.8\times 10^{-4}$ \\
    \hline
        \multirow{3}{*}{\blue{$e^- \mu^+ \to e^- \tau^+ a$}} & \multirow{3}{*}{$C^A_{\mu\tau} =C^V_{\mu\tau}$} & 1 & 1.4 \\
         & & 10 & 0.93\\
         & & 100 & $8.6\times10^{-2}$ \\
    \hline
        \multirow{3}{*}{\blue{$e^- \mu^+ \to \tau^- \mu^+ a$}} & \multirow{3}{*}{$C^A_{e\tau} =C^V_{e\tau}$} & 1 & $3.3 \times 10^{-2}$ \\
         & & 10 & $1.7 \times 10^{-2}$\\
         & & 100 & $2.3 \times 10^{-3}$ \\
    \hline\hline        
         \multirow{3}{*}{$\mu^+ \mu^+ \to \mu^+ e^+ a$} & \multirow{3}{*}{$C^A_{e\mu} = C^V_{e\mu}$} & 1 & $8.7 \times 10^{-4}$\\
         & & 10 & $5.1 \times 10^{-4}$\\
         & & 100 & $1.8 \times 10^{-4}$ \\
    \hline
        \multirow{3}{*}{\blue{$\mu^+ \mu^+ \to \mu^+ \tau^+ a$}} & \multirow{3}{*}{$C^A_{\mu\tau} = C^V_{\mu\tau}$} & 1 & $0.23$\\
         & & 10 & $0.14$\\
         & & 100 & $0.05$\\
    \hline\hline
    \end{tabular}
    \caption{  Same as Table~\ref{tab:process_2fa} for ALP production processes that only involve one LFV ALP vertex, hence the cross sections scale as $(100~\text{GeV}/f_a)^2$. The resulting cross sections are the same if the ALP couples to leptons with $C^A_{\ell_i\ell_j} = -C^V_{\ell_i\ell_j}$, while they are a factor of two smaller if either $C^A_{\ell_i\ell_j} = 0$ or $C^V_{\ell_i\ell_j} =0$.
    The most promising processes (highlighted in blue) are discussed in detail in the next sections.
    }
    \label{tab:process_1fa}
\end{table}

We now turn to the more promising class of processes featuring an on-shell ALP in the final state. Since the dominant diagrams require a single ALP interaction, such processes are affected by a milder suppression, $\sim 1/f_a^2$, than those considered in the previous subsection.
Table~\ref{tab:process_1fa} lists on-shell ALP production processes involving electroweak $Z/\gamma$ boson interactions and the LFV ALP couplings. The collision of $e^-$ and $\mu^+$ beams can yield the $2\to 2$ processes $e^-\mu^+\to a \gamma$ and $e^-\mu^+\to a Z$ through the $C^{A,V}_{e\mu}$ interaction and a photon/$Z$ boson radiation. The others are all $2\to 3$ processes from $t$-channel $Z/\gamma$ diagrams with an ALP emitted from an initial- or final-state lepton. Notice that those with a $\tau$ lepton in the final state are induced by the $C^{A,V}_{\ell\tau}$ couplings and their cross sections are enhanced by the $\tau$ mass, that is, by a factor $\sim (m_\tau/m_\mu)^2$\,---\,see  Eq.~\eqref{eq:gVA}\,---\,compared to the processes that only involve electrons and muons. 
Furthermore, processes with the ALP emitted from the muon have a substantially larger cross section than those with the ALP emitted from the electron, due to the large asymmetry of the beam energies in $e^-\mu^+$ collisions.
The highlighted processes have large enough cross sections to be a realistic target at $\mu$TRISTAN.\footnote{We do not study $e^-\mu^+\to a Z$, although its cross section can be even larger than $e^-\mu^+\to a \gamma$. In fact, the number of signal events would decrease significantly in exclusive searches targeting specific $Z$ decay modes. Furthermore, as we will see, the final-state particles produced in asymmetric $e^-\mu^+$ collisions tend to have a large pseudorapidity. This  substantially reduces the reconstruction efficiency of $Z$ bosons, as a large fraction of the decay products is outside the geometric acceptance of the detector.}
Hence, in the following analysis, we will focus on the following ALP production modes as a way to probe the three LFV ALP couplings: 
\begin{itemize}
    \item $C^{A,V}_{e\mu}$: $e^- \mu^+ \to a \gamma$\;; 
    \item $C^{A,V}_{\mu\tau}$: $e^- \mu^+ \to e^- \tau^+ a$ and $\mu^+ \mu^+ \to \mu^+ \tau^+ a$\;;
    \item $C^{A,V}_{e\tau}$: $e^- \mu^+ \to \tau^- \mu^+ a$\;.
\end{itemize}

\section{$\mu$TRISTAN sensitivity to LFV ALPs}
\label{sec:sensitivity}

In the following, we assess the $\mu$TRISTAN potential to test LFV ALP production through the processes outlined at the end of the previous section. 
Signal and background events are generated by means of \textsc{Madgraph}~\cite{Alwall:2011uj,Maltoni:2002qb}, \textsc{Pythia}~\cite{Sjostrand:2014zea}, and \textsc{Madspin}~\cite{Artoisenet:2012st}, while the detector response is simulated by \textsc{Delphes}~\cite{deFavereau:2013fsa}. The resulting events are analysed using \textsc{Root}~\cite{Brun:1997pa}. 
For both signal and background events, we apply the default basic generator-level cuts on the final-state photons or charged leptons in the \textsc{Madgraph} package, with the exception of their maximum pseudorapidity, for which we consider different options, as discussed in this section. 

We employ the following definition of statistical significance
\begin{equation}
\mathcal{S} = \frac{N_{\rm S}}{\sqrt{N_{\rm S}+N_{\rm B}}}\;,~~~N_{\rm S (B)}=\sigma_{\rm S (B)}\times \varepsilon_{\rm S (B)}\times \mathcal{L}\;,
\label{eqn:significance}
\end{equation}
where $N_{\rm S}$ ($\varepsilon_{\rm S}$) and $N_{\rm B}$ ($\varepsilon_{\rm B}$) are the event numbers (efficiencies) of signal and backgrounds, respectively. The production cross section $\sigma_{\rm S}$ ($\sigma_{\rm B}$) is multiplied by the branching fraction of the ALP (SM gauge bosons) decay processes relevant to the signal selection.
For later convenience, we define the signal efficiency as $\varepsilon_\text{S} \equiv \varepsilon_{\rm sel}\times \varepsilon_{\rm cut}$, that is, the product of the selection efficiency and the signal acceptance after cuts. The selection efficiency, accounting for the geometric acceptance of the detector and the particle identification probabilities, can be estimated counting the number of signal events remaining after the fast detector simulation performed by \textsc{Delphes}.
As we will discuss, several searches we consider are affected by a reduced $\varepsilon_{\rm sel}$, because the decay products of the ALP have a sizeable probability to be outside the geometric acceptance of the detector due to the large asymmetry of the electron and muon beams.

In the following, we are showing the expected exclusion potential of $\mu$TRISTAN at the 95\%~confidence level~(CL), corresponding to $\mathcal{S} =2$ in Eq.~\eqref{eqn:significance}, assuming that an integrated luminosity of $\mathcal{L}=10~\text{ab}^{-1}$ can be achieved. However, the most constraining searches being practically free of SM background, one can easily rescale the limits on the couplings we present as $\sim\left(10~\text{ab}^{-1}/\mathcal{L}\right)^{1/2}$.


\subsection{ALP production through $e^- \mu^+\to a \gamma$}
\label{sec:mueLFV}

We first analyse the process $e^- \mu^+\to a \gamma$ that is sensitive to LFV couplings of the ALP with electrons and muons. The analytical expression of the differential cross section is 
\begin{align}
    \frac{\dd \sigma}{\dd t}(e^- \mu^+\to a \gamma)\simeq \frac{\alpha}{16\pi}\frac{m_\mu^2}{f_a^2 }\frac{|C^A_{e\mu}|^2+ |C^V_{e\mu}|^2 }{t (m_a^2-t-s)} \simeq
    \frac{\alpha}{4\pi}\frac{|g^A_{e\mu}|^2+ |g^V_{e\mu}|^2 }{t (m_a^2-t-s)} \,,
\end{align}
where $\alpha$ is the fine-structure and constant and $t=(p_\mu-p_\gamma)^2$ in the limit $m_e,m_\mu\ll \sqrt{s}$. 
After integrating the above expression, we obtain the total cross section that, for $m_a\ll \sqrt{s}$, reads
\begin{align}   
\sigma(e^-\mu^+ \to a\gamma)\simeq \alpha \frac{m_{\mu}^2}{ f_a^2}\frac{ |C^A_{e\mu}|^2+ |C^V_{e\mu}|^2}{4s}|\eta|_\text{max}
\simeq \frac{\alpha}{s} \left( |g^A_{e\mu}|^2+ |g^V_{e\mu}|^2\right)|\eta|_\text{max}
\,,
\label{eq:cross-sec-emu-alpgamma}
\end{align}
where $|\eta|_\text{max}$ is the maximum pseudorapidity within the geometrical acceptance of the detector. For $|\eta|_\text{max} =2.5$, the above formula agrees with the \textsc{Madgraph} result reported in Table~\ref{tab:process_1fa} within 10\%.
As one can see, a larger cross section can be achieved by decreasing the energy of the muon beam $E_\mu$ and thus $\sqrt{s}$. However, the $\mu$TRISTAN luminosity is expected to decrease for lower values of $\sqrt{s}$~\cite{Hamada:2022mua}. Here, we consider the beam energy as in the $\mu$TRISTAN proposal, i.e.~$E_e=30$~GeV and $E_\mu=1$~TeV, while we comment on the impact on the sensitivity of different values of $\sqrt{s}$ in the concluding Section~\ref{sec:concl}.

The experimental signatures that could be used to search for this production process depend on the possible ALP decay modes, hence on its mass and couplings. 
Consequently, there could be different SM backgrounds which may or may not be trivially reducible. We consider several possible signals and backgrounds, distinguishing a ``heavy ALP'' scenario (with $m_a > m_e+m_\mu$), for which the ALP can decay as $a\to e\mu$ through the same LFV interaction as the production process, and a ``light ALP" case (with $m_a < m_e+m_\mu$), for which such process is kinematically closed. We start discussing the latter case.

\begin{figure}[t!]
    \centering
    \includegraphics[width=0.55\textwidth]{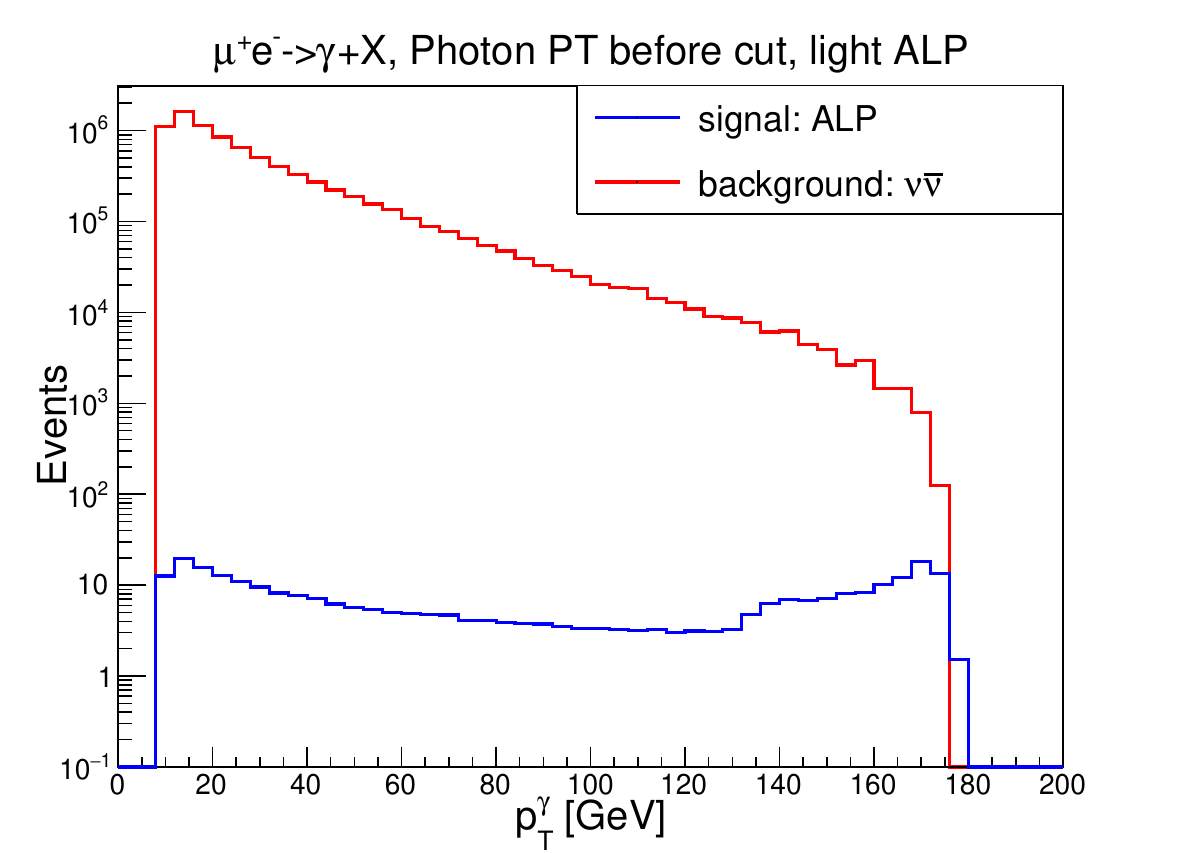}
\caption{  Distribution of the photon transverse momentum, $p_T^{\gamma}$, for $e^-\mu^+ \to \gamma +$~invisible events before cuts.
The blue line denotes the signal from production of a light invisible ALP, with $C^A_{e\mu}/f_a = C^V_{e\mu}/f_a= 0.01~\text{GeV}^{-1}$, corresponding to $|g^A_{e\mu}| \simeq |g^V_{e\mu}| \simeq 5.3\times 10^{-4}$. The red line is the SM background $e^-\mu^+ \to \gamma \nu_e\bar\nu_\mu$. The number of events is calculated including both production cross sections and assuming for the integrated luminosity $\mathcal{L} = 10~\text{ab}^{-1}$.
}
\label{fig:pt_met_emu_agamma_light}
\end{figure}

\paragraph{Light ALP.}
As discussed in Section~\ref{sec:LFV-ALP}, if $m_a<m_e+m_\mu$ and the LFC couplings are absent or suppressed (in particular, if $C_{ee}^A=0$) the ALP is long-lived and the signal events will feature a single photon and missing energy. The SM background is then given by the process $e^- \mu^+\to \gamma \nu_e \bar{\nu}_\mu$ that stems from exchange of a $W$ boson in the $t$ channel and a photon from initial state radiation or $W$ bosons' annihilation.
The cross-section is rather large, 840~fb, once the standard \textsc{Madgraph} generator-level cuts are applied. 

In Figure~\ref{fig:pt_met_emu_agamma_light}, we show the signal and BG distributions of the photon transverse momentum $p_T^\gamma$, which is the same as the transverse missing energy (MET). The distribution for the signal exhibits a peak around $p_T^\gamma \simeq \sqrt{s}/2 \simeq 170$~GeV from the Lorentz boost factor in the two-body final state. Instead, the $p_T^\gamma$ distribution of the background decreases exponentially with increasing energy. Therefore, we choose the kinematic cut $p_T^\gamma>170$~GeV, in order to distinguish the signal from the background.

If $C_{ee}^A\neq 0$ and the decay $a\to e^+e^-$ is kinematically allowed, this process occurs with nearly $100\%$ branching ratio, since the decay into photons is loop-suppressed (see Appendix~\ref{app:gamma}). The signature then depends on the ALP decay length shown in Figure~\ref{fig:alplifetime1}. Long-lived enough ALPs can be still searched for in events with a single photon and MET
but we expect a reduction in the sensitivity due to the fraction of ALPs decaying inside the detector. ALPs with a shorter lifetime can be sought in $e^-\mu^+ \to \gamma e^+ e^-$ events. In this case, the SM background is given by the process $e^-\mu^+\to \gamma e^- \bar{\nu}_\mu W^+ (\to e^+ \nu_e)$, whose cross section is $0.34$~fb. Because of the presence of neutrinos in the final state, a cut on the MET of the events can reduce this background significantly. In this case, we impose MET$<10$~GeV. 

\paragraph{Heavy ALP.}
If $m_a >m_e + m_\mu$, the ALP can decay as $a\to e^\pm\mu^\mp$ through the very same interaction involved in its production. In order to avoid any major SM background, one can select events with negative muons and positive electrons in the final state: $e^- \mu^+ \to \gamma a \to \gamma \mu^- e^+$. In the presence of LFV couplings only, one has $\text{BR}(a\to \mu^- e^+)=50\%$.
The process $e^-\mu^+\to \gamma \nu_e \bar \nu_\mu W^-(\to \mu^- \bar \nu_\mu)W^+(\to e^+ \nu_e)$ can in principle produce a reducible background. However, it can be completely neglected because of its tiny cross section, $\sigma_\text{B} \simeq 2 \times 10^{-5}$~fb.\footnote{ Strictly speaking, box diagrams with neutrino and $W$ propagators could induce $e^-\mu^+ \to \mu^- e^+ (\gamma)$ in the SM. However, such process is suppressed to negligible levels by the tiny neutrino masses, as it is always the case for LFV processes within the SM. In this case the suppression is of the order $\left({m_\nu}/{m_W}\right)^8 \approx 10^{-98}$!}

Additionally, if LFC ALP couplings exist, one can also search for $e^- \mu^+ \to a~\gamma$ followed by $a \to \mu^+ \mu^-$. If $C_{\mu\mu}^A=C_{e\mu}^A=C_{e\mu}^V$, the ALP decay branching fractions are ${\rm BR}(a\to \mu^+\mu^-) \simeq 50\%$ and ${\rm BR}(a\to e^+\mu^-) \simeq 25\%$ for $m_a \gg 2\, m_\mu$. The main SM background in this case is given by the process $e^-\mu^+\to \gamma \nu_e\bar{\nu}_\mu Z/\gamma^\ast (\to \mu^+\mu^-)$, whose cross section is 0.12~fb. 
We find that a missing energy cut, MET$<$15~GeV, is sufficient to eliminate this SM background.

\begin{table}[t!]
    \centering
    \renewcommand{\arraystretch}{1.4}
    \resizebox{\textwidth}{!}{
    \begin{tabular}{cccccccc}
    \hline\hline 
   \multirow{2}{*}{ALP couplings} & \multirow{2}{*}{$m_a$ (GeV)} &\multirow{2}{*}{Signature~(BR)} &  \multirow{2}{*}{SM background} &\multirow{2}{*}{Cut} & \multirow{2}{*}{$\sigma_{\text{B}}(\rm{fb})\times \varepsilon_{\rm B}$} & \multirow{2}{*}{$\varepsilon_{\text{S}} = \varepsilon_{\rm sel}\times \varepsilon_{\rm cut}$} & $2\sigma$ limit on $|g_{e\mu}^{V,A}|$ \\ 
   &  &  &  & & &  &($\mathcal{L} = 10~ \text{ab}^{-1}$)\\ \hline\hline
      \multirow{3}{*}{$C_{e\mu}^V=C_{e\mu}^A$} & \multirow{1}{*}{0.1} & \multirow{1}{*}{invisible~(100\%)} & \multirow{1}{*}{$e^- \mu^+ \to \gamma \nu_e \bar\nu_\mu$}  & \multirow{1}{*}{$p_T^\gamma > 170$ GeV} & \multirow{1}{*}{$840\times 0.004\%$} & \multirow{1}{*}{$ 95\%\times 8.3\%$}  & \multirow{1}{*}{$1.2\times 10^{-3}$} \\    
     & 10 & \multirow{2}{*}{$a\to e^+\mu^-$~(50\%)} &  \multirow{2}{*}{negligible} & \multirow{2}{*}{--} & \multirow{2}{*}{--} &  $14\%\times 100\%$  &  $2.2\times 10^{-4}$\\ 
     & 100 & & & & & $14\%\times 100\%$ & $2.1\times 10^{-4}$\\ \hline 
     \multirow{3}{*}{$C_{e\mu}^{V,A}=C^A_{\ell_i\ell_i}$} & \multirow{1}{*}{0.1}& \multirow{1}{*}{$a\to e^+ e^-$}~(100\%)$\,^\dag$ & \multirow{1}{*}{$e^-\mu^+\to \gamma e^- \bar{\nu}_\mu W^+ (\to e^+\nu_e)$} & \multirow{1}{*}{MET~$<$~10~GeV}& \multirow{1}{*}{$0.34\times6\%$}  &  \multirow{1}{*}{$14\%\times 73\%$} & \multirow{1}{*}{$8.4\times 10^{-4}$}\\ 
     & \multirow{2}{*}{10}& $a\to e^+\mu^-$~(25\%) & negligible & -- & -- &  $14\%\times 100\%$ &  $3.1\times10^{-4}$\\ 
      &&$a\to \mu^+\mu^-$~(50\%) & $e^-\mu^+\to \gamma \nu_e\bar{\nu}_\mu Z/\gamma^\ast(\to\mu^+\mu^-)$ & MET~$<$~15~GeV & $0.12 \times 5\%$ &  $39\%\times42\%$  & $5.9\times10^{-4}$ \\
        \hline\hline
    \end{tabular}}
    \caption{  Summary of the possible searches for ALPs produced in $e^-\mu^+ \to a\gamma$ for different ALP masses and couplings. The upper block shows the results for exclusively LFV ALPs, the lower block for ALPs also with LFC couplings. The third column reports the branching ratio (BR) of the ALP decay under consideration\,---\,the process denoted by $^\dag$ has $\text{BR} \simeq 100\%$ but only about 40\% of the decays occur in the inner detector for the chosen value of $m_a$. 
    The applied cuts (if any), the BG cross section $\sigma_\text{B}$ and efficiency $\varepsilon_\text{B}$, the signal efficiency 
    $\varepsilon_{\text{S}} \equiv \varepsilon_{\text{sel}}\times\varepsilon_{\text{cut}}$ (see the main text for details) are shown in the fifth, sixth and seventh columns, respectively. In the last column, we report the resulting $2\sigma$ limit on $|g_{e\mu}^{V,A}| \simeq m_\mu |C_{e\mu}^{V,A}|/2f_a$ (with $C_{e\mu}^{V}=C_{e\mu}^{A}$) for $\mathcal{L}=10~\text{ab}^{-1}$. For the BG-free searches these limits scale as $\sim (10~\text{ab}^{-1}/\mathcal{L})^{1/2}$.
    }
    \label{tab:emu_agamma_cutresult}
\end{table}

\paragraph{$\mu$TRISTAN sensitivity.} 
In Table~\ref{tab:emu_agamma_cutresult}, we summarise signals and backgrounds for the different ALP masses and coupling scenarios discussed above. We also show the kinematic cuts that effectively reduce the backgrounds in case they are present, and then compute the background and signal cut efficiencies. In order to estimate the signal significance, we additionally include the effect of the particle identification capabaility and geometric acceptance of the detector, $\varepsilon_{\rm sel}$, assuming as a requirement for the pseudorapidity of photons and leptons $|\eta| < 2.5$. We finally obtain a $2\sigma$ lower bound on the flavour violating coupling $g_{e\mu}^{V,A}$ for an integrated luminosity of 10~ab$^{-1}$ using Poisson statistics as in Eq.~\eqref{eqn:significance}.
We find that $\mu$TRISTAN has the capability to constrain LFV ALP couplings as low as $|g_{e\mu}^{V,A}|\sim 10^{-3}-10^{-4}$. 

\begin{figure}[t!]
    \centering
    \includegraphics[width=0.55\textwidth]{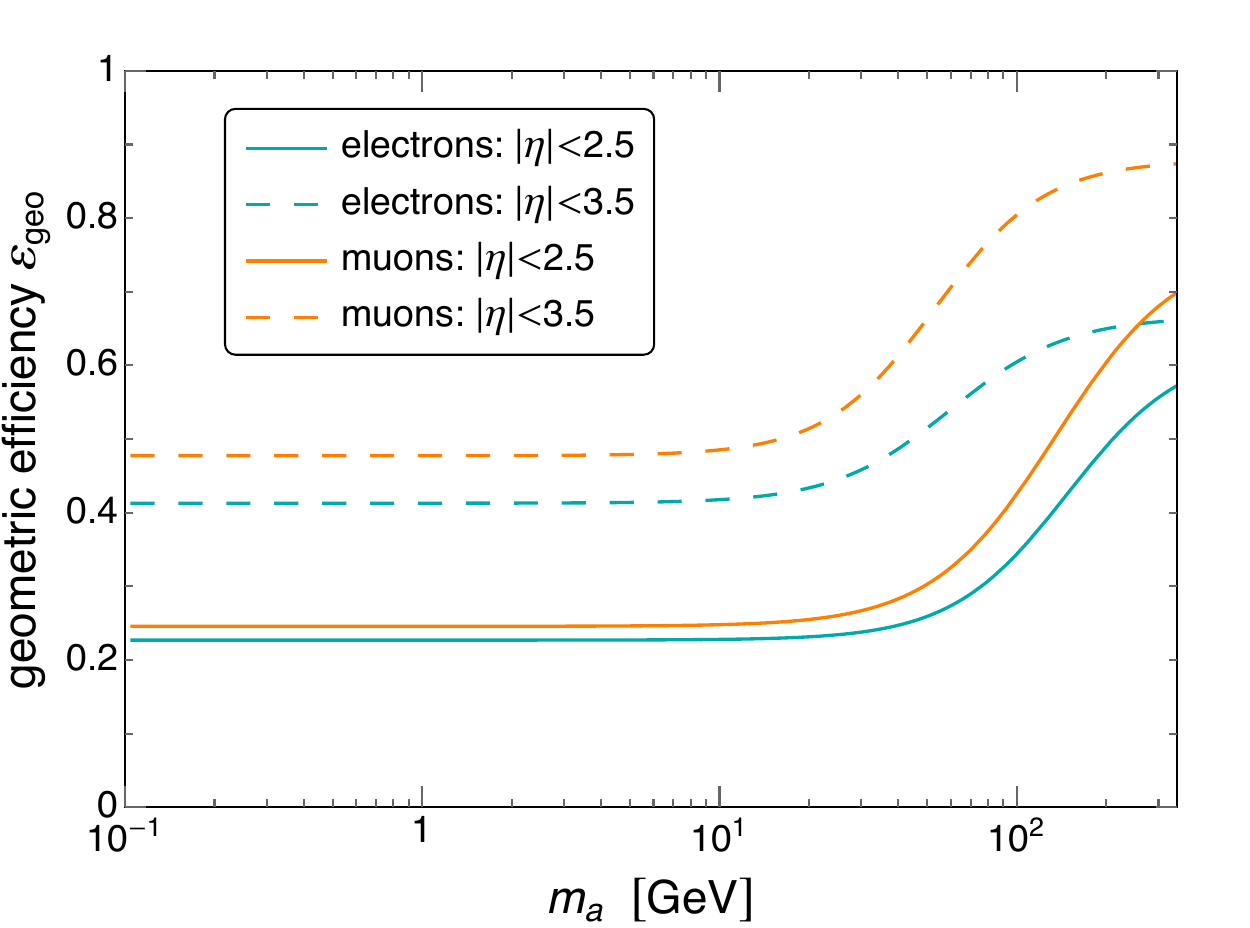}
    \caption{Geometric efficiency of electrons and muons from ALP decays as a function of the ALP mass, for two different designs of the detector: $|\eta|_{\text{max}} = 2.5$ (solid lines) and $|\eta|_{\text{max}} = 3.5$ (dashed lines).}
    \label{fig:emu_eta_acceptance}
\end{figure}

As mentioned before, because of the planned asymmetric beam configuration of $\mu$TRISTAN, the ALP decay products tend to fly along the muon beam direction, with a very large pseudorapidity $|\eta|$.\footnote{This partly occurs also for Higgs decays~\cite{Hamada:2022mua}, which are the main target of the $\mu$TRISTAN proposal.} In order to study the impact on the sensitivity of this effect, we consider two possible detectors with different values of $|\eta|_\text{max}$: one with a standard geometric acceptance $|\eta|<2.5$, and one more suitable for the $\mu$TRISTAN design with $|\eta|<3.5$. In Figure~\ref{fig:emu_eta_acceptance}, we show the probability\,---\,that we call ``geometric efficiency'' $\varepsilon_\text{geo}$\,---\,for muons and electrons from ALP decays to be within the geometric acceptance of the detector, ignoring particle identification. We plot $\varepsilon_\text{geo}$ as a function of the ALP mass, showing how  it varies in the case of a detector with $|\eta|_\text{max} =3.5$ (dashed lines) instead of $|\eta|_\text{max} =2.5$ (solid lines). As expected, $\varepsilon_{\rm geo}$ increases if a larger lepton pseudorapidity range is allowed. However, even in such a case, a substantial proportion of signal events is lost, especially for light ALPs ($50\%-60\%$).\footnote{We checked that the presence of a forward muon detector covering large pseudorapidities (up to $|\eta|_\text{max} = 8$) does not substantially improve the reach of the most sensitive search we consider\,---\,the one targeting the LFV ALP decay $a\to e^+\mu^-$\,---\,which is still limited by the reduced probability of detecting the positron. We also checked that, instead, the angular separation between positron and photon does not give any further limitation on the detection of our signal. For instance, imposing $\theta_{e\gamma}>0.2$ has no appreciable impact on our estimated sensitivities.}

\begin{figure}[t!]
    \centering
    \subfigure[LFV ($e$-$\mu$) couplings only]{\includegraphics[width=0.48\textwidth]{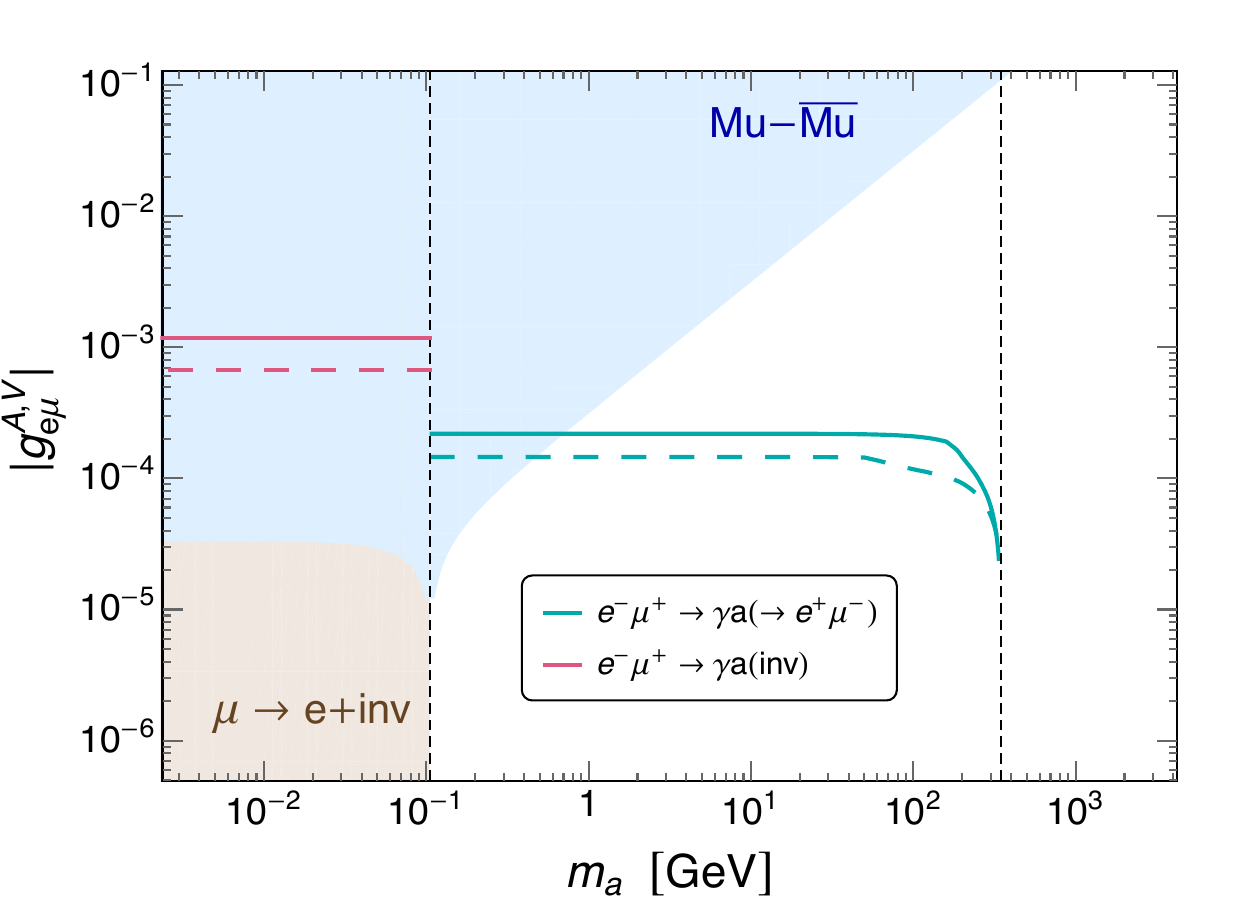}}
    \hfill
    \subfigure[LFV ($e$-$\mu$) and LFC couplings]{\includegraphics[width=0.48\textwidth]{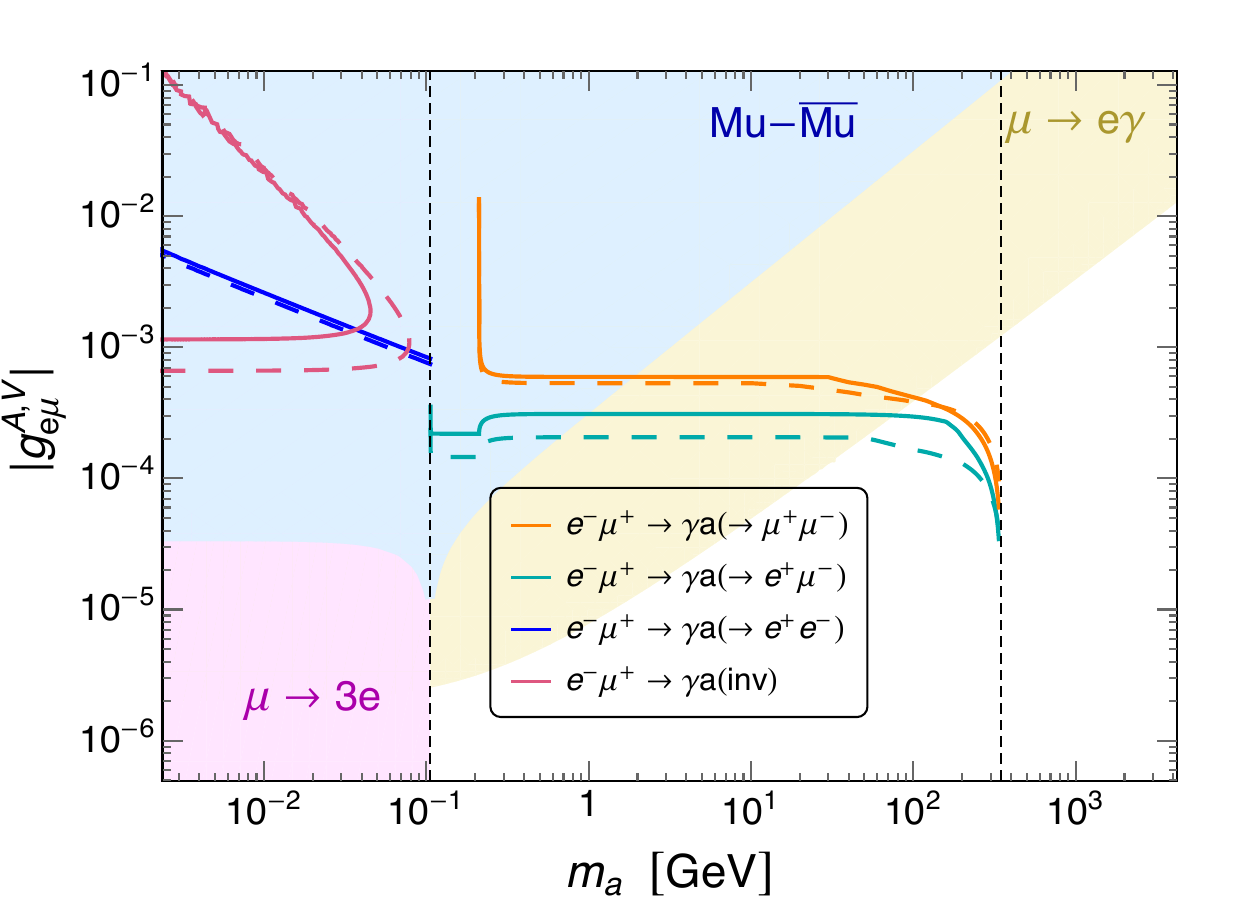}}
    \caption{Sensitivity of $\mu$TRISTAN to ALPs from $e^- \mu^+\to a \gamma$. The expected 95\% CL exclusion limits on $|g^{V,A}_{e\mu}|$ as a function of $m_a$ are shown for a purely LFV ALP with $C^A_{e\mu}=C^V_{e\mu}$ (left panel) and an ALP with LFC and LFV couplings and $C^A_{ee}=C^A_{\mu\mu}=C^{A,V}_{e\mu}$ (right panel). Solid (dashed) lines denote sensitivities of a detector with $|\eta|_{\text{max}} = 2.5$ ($|\eta|_{\text{max}} = 3.5$). The limits scale as $\sim(10~\text{ab}^{-1}/\mathcal{L})^{1/2}$. The two vertical dashed lines indicate the $m_\mu + m_e$ threshold and the centre of mass energy of the $e^-\mu^+$ collisions respectively. The coloured regions are excluded by searches for low-energy LFV processes: muonium-antimuonium oscillations (blue), two-body muon decay into an electron and an invisible boson (brown), $\mu \to eee$ (pink), $\mu\to e\gamma$ (yellow). See main text for details.}
\label{fig:constraint_gemu_zoom_to_collider}
\end{figure}

 In Figure~\ref{fig:constraint_gemu_zoom_to_collider}, we plot the resulting sensitivity of $\mu$TRISTAN in the $m_a-|g^{V,A}_{e\mu}|$ plane for a purely LFV ALP with $C_{e\mu}^{V}=C_{e\mu}^{A}$ (left panel) 
 and an ALP with all LFC and LFV couplings $C_{\ell_i\ell_j}^{A,V}$ equal (right panel). The solid (dashed) curves show the lower limits on the couplings that can be obtained by a detector with $|\eta|_\text{max} =2.5$ ($|\eta|_\text{max} =3.5$) searching for the different signatures discussed in this section, assuming an integrated luminosity of $\mathcal{L}=10~\text{ab}^{-1}$. As mentioned, these limits scale as $\sim\left(10~\text{ab}^{-1}/\mathcal{L}\right)^{1/2}$ in the situations where there are no background events. 

For a light ALP with LFC couplings (right plot), let us note the interplay between the search for an invisible ALP, which loses sensitivity if the ALP lifetime is reduced (for larger values of $m_a$ and larger couplings), and the search for $a\to e^+ e^-$ which is more sensitive to the latter situation, cf.~Figure~\ref{fig:alplifetime1}.
 
As we can see from both plots, the best sensitivity on the LFV couplings can be achieved by the background-free search for $a\to e^+ \mu^-$ if that is kinematically allowed. This search can test the parameter space substantially beyond the limits from low-energy LFV processes (shown as coloured regions) especially for a purely LFV ALP (left plot). For an ALP with also LFC couplings (right plot), $\mu$TRISTAN could test some unconstrained region only if $m_a \gtrsim 50$~GeV. In either scenario, low-energy constraints on light LFV ALPs already exclude the complete range of couplings that could be tested at $\mu$TRISTAN. In the following, we discuss in detail the current and future limits from low-energy LFV processes.

\begin{figure}[t!]
    \centering
    \quad
    \subfigure[LFV ($e$-$\mu$) couplings only]{\includegraphics[width=0.48\textwidth]{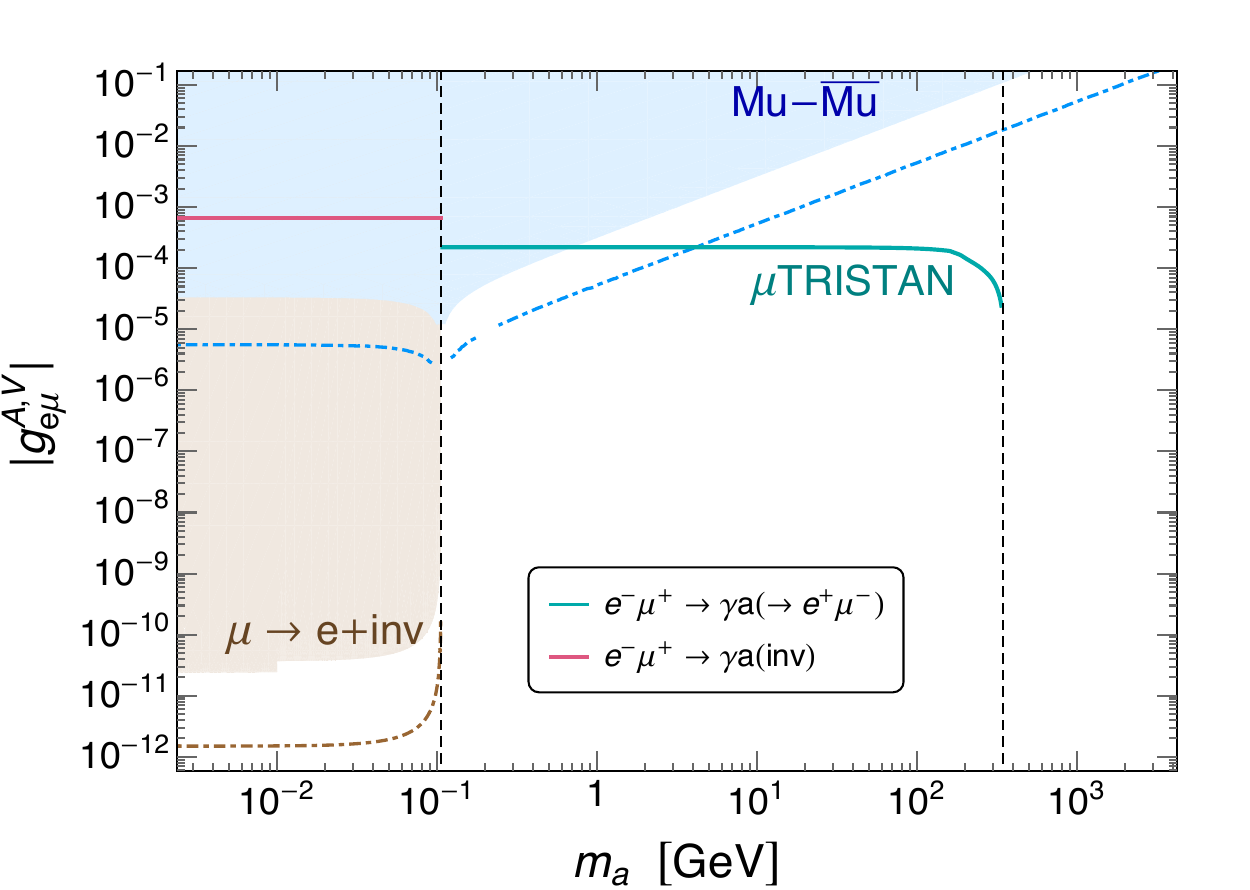}}
    \hfill
    \subfigure[LFV ($e$-$\mu$) and LFC couplings]{\includegraphics[width=0.48\textwidth]{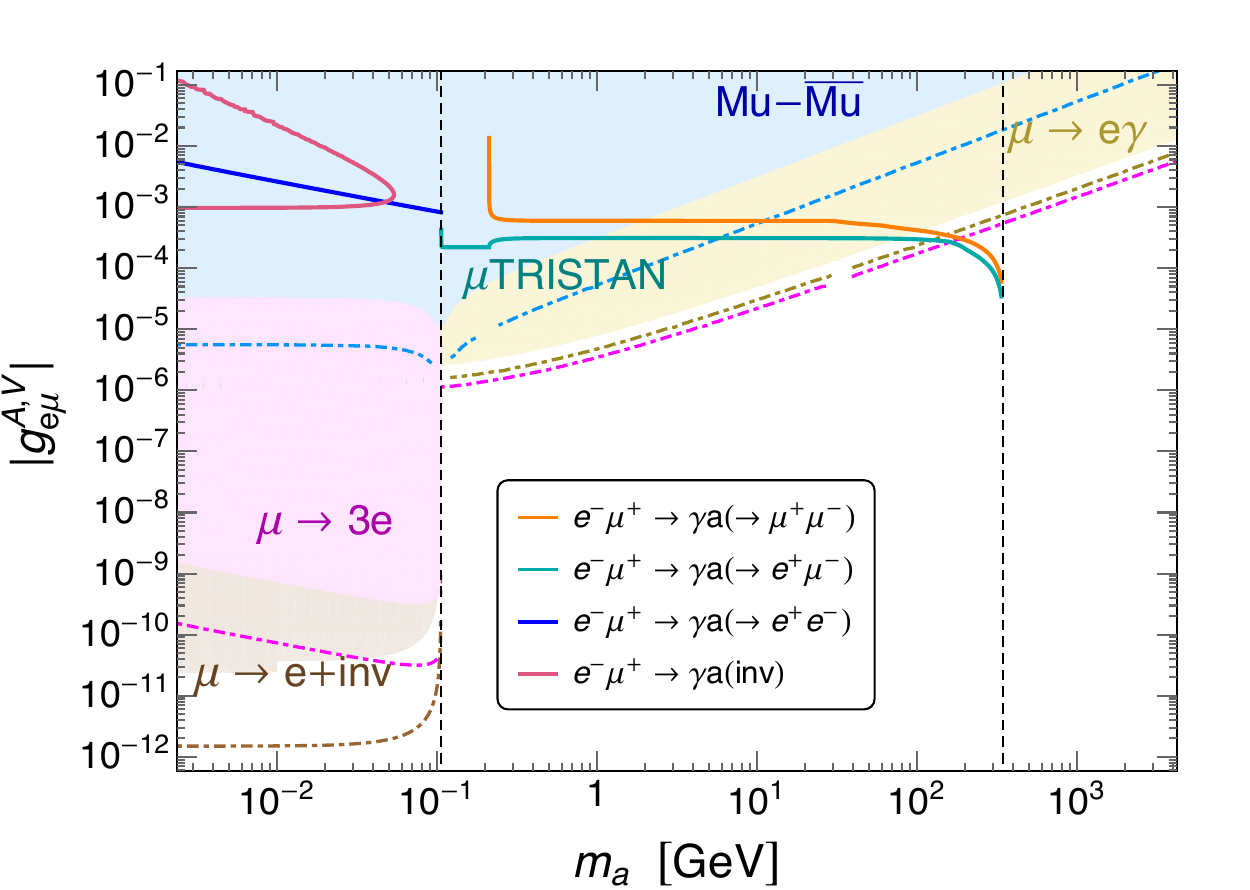}}
    \caption{Overview of the present and future constraints on LFV ALPs in the $m_a-|g^{V,A}_{e\mu}|$ plane and comparison with the $\mu$TRISTAN sensitivity as estimated in this work. The model's parameters and the colour coding are as in Figure~\ref{fig:constraint_gemu_zoom_to_collider}. The dot-dashed lines represent the expected sensitivities of running or future LFV experiments. See main text for details.}
    \label{fig:constraint_gemu}
\end{figure}

\paragraph{Present and future LFV and $g-2$ constraints.} 
In Figure~\ref{fig:constraint_gemu}, we summarise present and expected future constraints from searches for low-energy LFV processes, denoting them as coloured regions and dot-dashed lines, respectively. The ALP contributions to these processes are computed by means of the expressions reported in the Appendix~\ref{app:LFV}.

The most model-independent constraint on an ALP with LFV interactions with electrons and muons stems from mixing between muonium $\text{Mu}$ (that is, a $\mu^+e^-$ bound state) and antimuonium $\overline{\text{Mu}}$ ($\mu^-e^+$), since the ALP contribution to this process solely relies on the $C^{A,V}_{e\mu}$ couplings. 
The blue regions in Figures~\ref{fig:constraint_gemu_zoom_to_collider} and~\ref{fig:constraint_gemu} are excluded by the MACS experiment limit on the $\text{Mu}-\overline{\text{Mu}}$ oscillation probability, $P_{\text{Mu}\,\overline{\text{Mu}}}<8.3\times 10^{-11}$ at 90\%~CL~\cite{Willmann:1998gd}.
As we can see, this is the only extant constraint on a heavy ALP with the LFV couplings $C^{A,V}_{e\mu}$ only. In such a case, according to our estimate, 
$\mu$TRISTAN can test unconstrained regions of the parameter space if $m_a \gtrsim 1$~GeV.
The proposed MACE experiment is expected to improve the limit on muonium oscillations by three orders of magnitude, down to $P_{\text{Mu}\,\overline{\text{Mu}}}<7\times 10^{-14}$~\cite{Bai:2022sxq,mace_talk}. The MACE impact on our parameter space is shown as a blue-dashed line in Figure~\ref{fig:constraint_gemu}. Let us note the nice complementarity of MACE and $\mu$TRISTAN, with the former providing a better sensitivity for $m_a \lesssim 3-4$~GeV. 

If the heavy ALP also enjoys substantial LFC interactions with electrons or muons, the LFV decay $\mu\to e\gamma$ is induced by an ALP-lepton loop. As shown in the right plots of Figures~\ref{fig:constraint_gemu_zoom_to_collider} and~\ref{fig:constraint_gemu}, the current 90\%~CL limit set by the MEG experiment, $\text{BR}(\mu\to e \gamma) < 4.2\times 10^{-13}$~\cite{MEG:2016leq}, gives a more stringent constraint (yellow region) on our parameter space than $\text{Mu}-\overline{\text{Mu}}$ oscillations, if 
$C^A_{ee}=C^A_{\mu\mu}=C^{A,V}_{e\mu}$. As a consequence, $\mu$TRISTAN could be only sensitive to this kind of ALP models for $m_a$ in the $\mathcal{O}(100)$~GeV mass range.
The expected limit, $\text{BR}(\mu\to e \gamma) < 6\times 10^{-14}$, of the currently running MEG~II experiment~\cite{MEGII:2018kmf} is shown as a yellow dot-dashed line in Figure~\ref{fig:constraint_gemu}.

In the light ALP regime, LFV muon decays into an on-shell ALP provide by far the strongest constraints.
If the ALP is long-lived, the limit on $\text{BR}(\mu\to e a)$ with an invisible $a$ depends on the ALP mass and the chirality of its couplings~\cite{Calibbi:2020jvd}, ranging from $5.8\times 10^{-5}$~\cite{TWIST:2014ymv} for a (practically) massless boson coupling mainly to left-handed leptons (thus to a $V-A$ current) to $2.5\times 10^{-6}$~\cite{Jodidio:1986mz,Calibbi:2020jvd} if the couplings to right-handed leptons (hence to a $V+A$ current) dominate, such as in our benchmark scenario. For ALPs in the mass range $10~\text{MeV} \lesssim m_a \lesssim 90~\text{MeV}$, the dependence on $m_a$ is mild in the $V+A$ case and the average upper bound is $6\times 10^{-6}$~\cite{TWIST:2014ymv}, the limit that we employ here. The region excluded by searches for an invisible ALP in muon decays is shown in brown in Figure~\ref{fig:constraint_gemu}, while the dot-dashed brown line is the expected future limit of the Mu3e experiment, $\text{BR}(\mu\to e a) \lesssim 10^{-8}$~\cite{Perrevoort:2018ttp}.
In the presence of couplings to electrons, searches for $\mu\to e a(\to e^+e^-)$ can be also sensitive to our parameter space. The limit from $\mu\to e\gamma$ is stronger if the ALP is off-shell, while for on-shell short-lived ALPs the current limit $\text{BR}(\mu\to eee) < 1.0\times 10^{-12}$~\cite{Bellgardt:1987du} excludes the pink region of our figures. The Mu3e expected future bound, $\text{BR}(\mu\to eee) \lesssim 10^{-16}$~\cite{Blondel:2013ia,Perrevoort:2024qtc}, is shown as a dot-dashed pink line. In order to assess the relative importance of searches for invisible and visible ALPs, we used information on the ALP lifetime\,---\,illustrated in Figure~\ref{fig:alplifetime1}\,---\,as explained in Appendix~\ref{app:LFV}.  

LFV ALP interactions also contribute to both the electron and the muon anomalous magnetic moments through ALP-lepton loops. In the presence of LFC interactions, there are important additional contributions, especially due to the induced ALP-photon coupling~\cite{Marciano:2016yhf,Cornella:2019uxs}.
However, we find that limits on non-standard contributions to the electron and muon $g-2$ barely affect the region of the parameter space that can be tested by $\mu$TRISTAN beyond the limits from LFV observables, if no LFV coupling involving $\tau$ leptons is present. For this reason and in order to avoid an excess of information in the figures of this section, we do not show such limits here. More details and a discussion of the currently uncertain SM predictions of both observables are presented in Appendix~\ref{app:g-2}.


\subsection{ALP production through LFV $\tau$ interactions}
\label{sec:tauLFV}

\begin{figure}[t!]
\centering
{\includegraphics[width=0.49\textwidth]{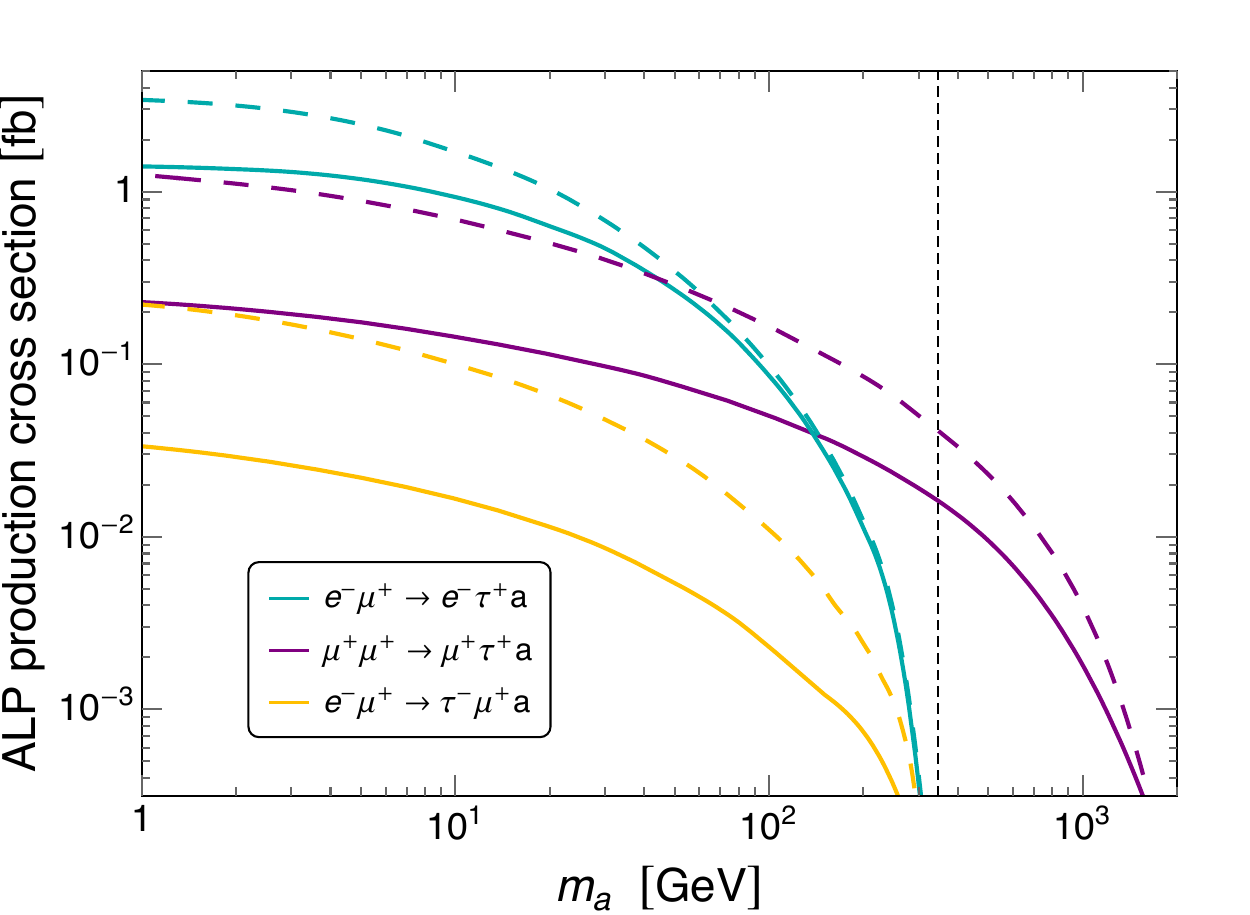}}
\hfill
{\includegraphics[width=0.49\textwidth]{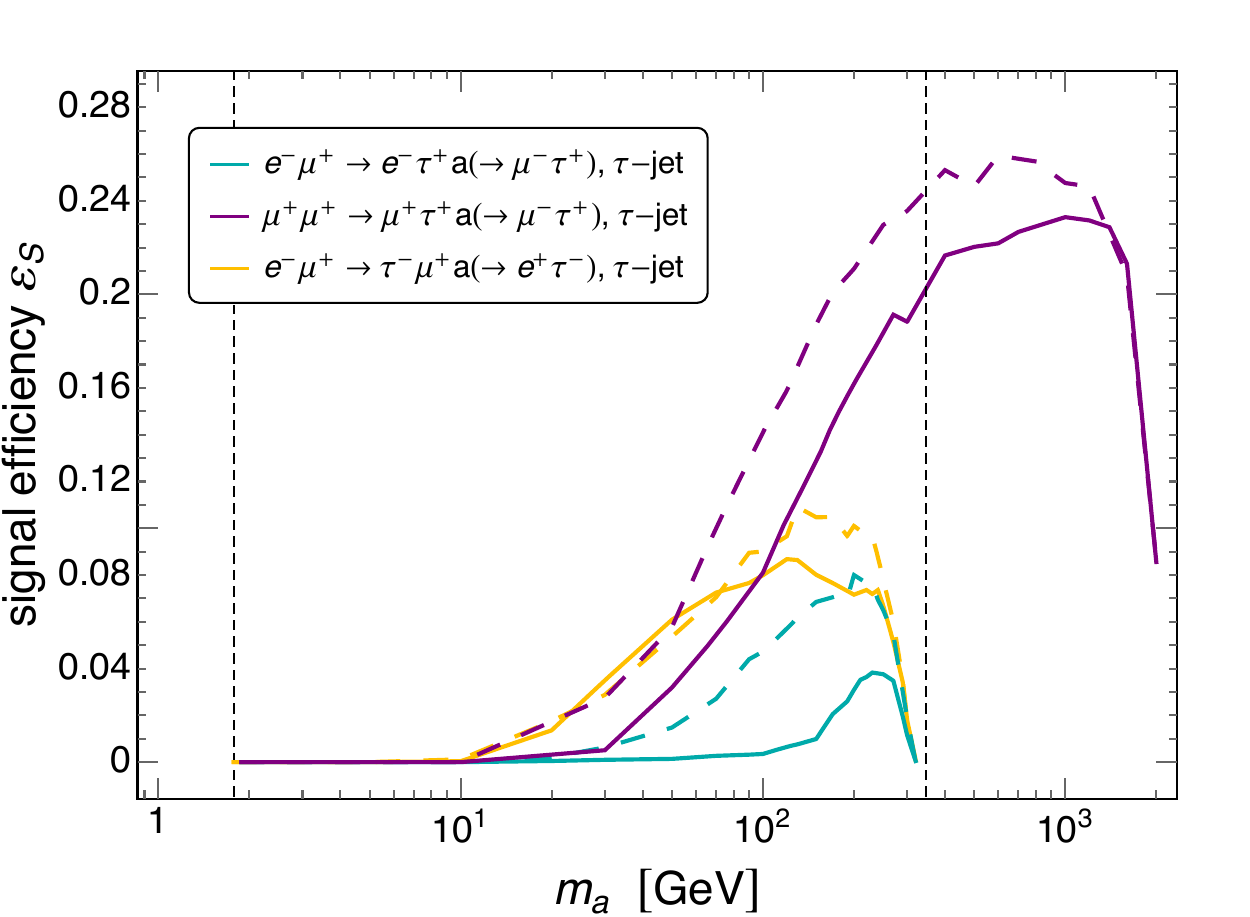}}
\caption{  
Left panel: total cross section of ALP production processes through LFV ALP-$\tau$ interactions at $\mu$TRISTAN for $|\eta|_{\text{max}} = 2.5$  (solid lines) and $|\eta|_{\text{max}} = 3.5$ (dashed lines), for $C^A_{\ell\tau}/f_a=C^V_{\ell\tau}/f_a = 0.01~\text{GeV}^{-1}$ ($\ell =e\,\text{or}\,\mu$), corresponding to $|g^A_{\ell\tau}| \simeq |g^V_{\ell\tau}| \simeq 8.9\times 10^{-3}$. 
Right panel: signal efficiency of searches for the  processes of the left plot with hadronic taus in the final state.
}
\label{fig:cs_taustate_ma}
\end{figure}

We now proceed to discuss the other promising search channels, which can target ALP production with an associated tau lepton in the final state as highlighted in Table~\ref{tab:process_1fa}. 
The cross sections of these processes as a function of the ALP mass for beams with energies as in the $\mu$TRISTAN proposal, $E_e = 30$~GeV and $E_\mu = 1$~TeV, are shown in the left plot of Figure~\ref{fig:cs_taustate_ma}. The solid lines correspond to a generator-level cut on the pseudorapidity of the final-state leptons (including the $\tau$) of $|\eta|<2.5$, the dashed lines to $|\eta|<3.5$.

For what concerns $\tau$ detection, one can use either the hadronic $\tau$ decays or the leptonic ones. In the former case, the resulting tau jets posses distinctive features. They are very collimated, exhibit a low particle multiplicity, mostly one or three prongs, and are associated to a relevant fraction of electromagnetic energy deposition mainly due to the decay into photons of neutral pions. Modern detectors are expected to yield an identification efficiency for hadronic taus as large as 90\%, see e.g.~Ref.~\cite{CEPCStudyGroup:2018ghi}. 

In the following, we examine several signatures that can be probed at $\mu$TRISTAN depending on the ALP mass and decay channels, as well as on the selection of hadronic or leptonic $\tau$ decays.

\begin{table}[b!]
    \centering
        \renewcommand{\arraystretch}{1.4}
    \begin{tabular}{cccc}
        \hline\hline
         ALP couplings & Signal & SM background &$\sigma_\text{B}$~(fb)  \\ \hline\hline
       \multirow{2}{*}{$C^{V,A}_{\mu\tau}$} 
        & $e^- \mu^+ \to e^- \tau^+ a_\text{inv}$ & $e^- \mu^+ \to e^- \tau^+ \bar{\nu}_\mu \nu_\tau$ & 24\\ 
        & $\mu^+ \mu^+ \to \mu^+ \tau^+ a_\text{inv}$& $\mu^+ \mu^+ \to \mu^+ \tau^+ \bar{\nu}_\mu \nu_\tau$ & 180 \\
        \hline
        $C^{V,A}_{e\tau}$ & $e^- \mu^+ \to \tau^- \mu^+ a_\text{inv}$ & $e^- \mu^+ \to \tau^- \mu^+ \bar{\nu}_\tau \nu_e$ & 4.8\\
        \hline\hline
    \end{tabular}
    \caption{SM background processes for a light ALP, $m_a< m_\tau + m_\ell$, appearing as missing energy ($a_\text{inv}$). The displayed $\sigma_\text{B}$ are computed requiring that all visible final-state particles have $|\eta|<2.5$.
    }
    \label{tab:alp_inv}
\end{table}
\begin{table}[b!]
    \centering
        \renewcommand{\arraystretch}{1.4}
        \resizebox{\textwidth}{!}{
    \begin{tabular}{cccc}
        \hline\hline
         ALP couplings & Signal & SM background &$\sigma_\text{B}$~(fb)  \\ \hline\hline
       \multirow{2}{*}{$C^{V,A}_{\mu\tau}=C^{A}_{\mu\mu}$} 
        & $e^- \mu^+ \to e^- \tau^+ a(\to \mu^+\mu^-)$  & $e^-\mu^+\to e^-  \bar{\nu}_\mu W^+(\to \tau^+\nu_\tau) Z/\gamma^* (\to\mu^+\mu^-)$  & $5.6\times 10^{-3}$ \\
        & $\mu^+ \mu^+ \to \mu^+ \tau^+ a(\to \mu^+\mu^-)$ &  $ \mu^+\mu^+\to \mu^+ \bar{\nu}_\mu W^+(\to \tau^+\nu_\tau) Z/\gamma^*(\to\mu^+\mu^-)  $  & 0.44 \\
        \hline
        $C^{V,A}_{e\tau}=C^{A}_{\mu\mu}$ & $e^- \mu^+ \to \tau^- \mu^+ a(\to \mu^+\mu^-)$ & $e^-\mu^+\to \nu_e \mu^+ W^-(\to \tau^-\bar\nu_\tau) Z/\gamma^*(\to\mu^+\mu^-)  $  & $7.6\times 10^{-5}$ \\
        \hline\hline
    \end{tabular}}
    \caption{SM background processes for a light ALP with $2m_\mu < m_a< m_\tau + m_\ell$ and LFC couplings to muons, such that the decay $a\to \mu^+\mu^-$ occurs inside the detector. The displayed $\sigma_\text{B}$ are computed requiring that all visible final-state particles have $|\eta|<2.5$. The illustrative production mode $Z/\gamma^\ast \to \mu^+\mu^-$ here and below actually includes all possible production processes of $\mu^+\mu^-$ in our background event generation.}
    \label{tab:alp_mumu}
\end{table}

\paragraph{Light ALP.}
We first consider ALPs whose decays into $\tau$ leptons are kinematically forbidden, that is, with $m_a < m_\tau+m_\ell$ (where $\ell =e$ or $\mu$). If the LFC interactions are absent, the ALP is long-lived and the processes in Table~\ref{tab:process_1fa} yield signatures involving missing energy. In such a case, the SM processes with neutrinos in the final state listed in Table~\ref{tab:alp_inv} look exactly like our ALP signals. As we can see, the BG cross sections are much larger than the ALP production cross sections shown in Figure~\ref{fig:cs_taustate_ma}. In addition, we found no kinematical distribution giving a handle to reduce the background.
Hence, the ALP emission signal induced by the LFV $C^{V,A}_{\ell\tau}$ couplings is completely overshadowed by the SM background and it is not possible to set a limit on such couplings from $\mu$TRISTAN in this case.

For ALPs coupling with muons, one could instead search for the decay $a \to \mu^+ \mu^-$, if it is kinematically allowed. The SM background processes for these modes are shown in Table~\ref{tab:alp_mumu}. As one can see, the associated $\sigma_\text{B}$ are not always negligible compared to our signals. Furthermore, as discussed in the following, hadronic $\tau$ tagging is not effective in the light ALP regime, such that one should rely on leptonic $\tau$ decays, further reducing the signal cross section and introducing more reducible SM background modes. Finally, similarly to the $\mu$-$e$ LFV case studied above, the light ALP regime is already strongly constrained by LFV processes. For these reasons, we refrain from a more quantitative study of this case and we move to discuss the heavy ALP regime.

\paragraph{Heavy ALP.}
In the rest of the section, we study the case  $m_a>m_\tau+m_\ell$, where the same LFV interactions involved in the ALP production processes are responsible for it to decay promptly as $a\to \ell\tau$.
As in the $\mu$-$e$ case, we select as signal the combination of charges of the final-state leptons that differs from the initial-state one: $a\to \mu^-\tau^+$ for the searches targeting ALP couplings to $\mu-\tau$, $a\to e^+\tau^-$ for the  $e-\tau$ ones. For simplicity, we do not consider the situation where $C^{A,V}_{\mu\tau}$ and $C^{A,V}_{e\tau}$ are simultaneously non-vanishing.
Notice that in the presence of LFC couplings with muons or taus, if $C^{A,V}_{\ell\tau} = C^{A}_{\mu\mu} = C^{A}_{\tau\tau}$ ($\ell =e\,\text{or}\,\mu$), $\text{BR}(a\to \tau^+\tau^-) \approx \text{BR}(a\to \ell^\pm\tau^\mp)$ and $\text{BR}(a\to \mu^+\mu^-) \approx (m_\mu / m_\tau)^2 \text{BR}(a\to \ell^\pm\tau^\mp)$, as one can see from Eq.~\eqref{eq:atoellell}. Hence, the decays into muons are irrelevant while those into $\tau$ leptons yield double signal events than the LFV decays, given that only one charge combination is selected for the latter ones. However, the $a\to\tau\tau$ signal is penalised by a reduced selection efficiency of hadronic taus\,---\,as we discuss below\,---\,or by the $\tau$ leptonic branching fractions. Hence, we only focus on LFV ALP decays in the following.

\begin{table}[b!]
    \centering
    \renewcommand{\arraystretch}{1.4}
    \resizebox{\textwidth}{!}{
    \begin{tabular}{cccc}
        \hline\hline
        ALP couplings & Signal & SM background& $\sigma_\text{B}$~(fb)  \\ \hline\hline
      \multirow{2}{*}{$C^{V,A}_{\mu\tau}$} & 
         $e^- \mu^+ \to e^- \tau_h^+ a(\to \tau_h^+ \mu^-)$ &  $e^-\mu^+ \to e^- \bar{\nu}_\mu W^+(\to\tau_h^+\nu_\tau) W^+(\to\tau_h^+\nu_\tau) W^-(\to\mu^-\bar\nu_\mu)$ & $2.1\times10^{-8}$ \\ 
      & $\mu^+ \mu^+ \to \mu^+ \tau_h^+ a(\to\tau_h^+ \mu^-)$ &  $\mu^+ \mu^+ \to \bar\nu_\mu \bar \nu_\mu W^+(\to \tau_h^+ \nu_\tau) W^+(\to \tau_h^+ \nu_\tau) Z/\gamma^*(\to\mu^+\mu^-)$ & $1.4\times 10^{-3}$ \\
      \hline
         $C^{V,A}_{e\tau}$ & $e^- \mu^+ \to \tau_h^- \mu^+ a(\to \tau_h^- e^+)$ & $e^-\mu^+ \to \nu_e \mu^+  W^+(\to e^+\nu_{e})W^-(\to\tau_h^-\bar\nu_{\tau})W^-(\to\tau_h^-\bar\nu_{\tau})$ & $1.1\times10^{-8}$ \\
        \hline\hline
    \end{tabular}}
\caption{
SM background processes for a heavy ALP, $m_a > m_\tau + m_\ell$, with hadronically decaying $\tau$ leptons  (denoted as $\tau_h$) in the final state. The displayed $\sigma_\text{B}$ are computed requiring that all visible final-state particles have $|\eta|<2.5$.}
\label{tab:alp_hadtaus}
\end{table}
\begin{table}[b!]
    \centering
        \renewcommand{\arraystretch}{1.4}
        \resizebox{\textwidth}{!}{
    \begin{tabular}{cccc}
        \hline\hline
         ALP couplings & Signal & SM background &$\sigma_\text{B}$~(fb)  \\ \hline\hline
       \multirow{2}{*}{$C^{V,A}_{\mu\tau}=C^{A}_{\mu\mu}$} 
        & $e^- \mu^+ \to e^- \tau_e^+ a(\to \tau_e^+\mu^-)$ & $e^-\mu^+\to \nu_e\bar{\nu}_\mu W^-(\to \mu^-\bar\nu_\mu) W^+(\to e^+\nu_e) Z/\gamma^*(\to e^+ e^-) $ & $2.3\times10^{-7}$ \\ 
        & $\mu^+ \mu^+ \to \mu^+ \tau_e^+ a(\to \tau_e^+\mu^-)$ & $\mu^+\mu^+\to  \bar\nu_\mu\bar{\nu}_\mu W^+(\to e^+\nu_e) W^+(\to e^+\nu_e) Z/\gamma^*(\to\mu^+\mu^-)$  &  $3.3\times10^{-3}$ \\
        \hline
        $C^{V,A}_{e\tau}=C^{A}_{\mu\mu}$ & $e^- \mu^+ \to \tau_\mu^- \mu^+ a(\to \tau_e^+\mu^-)$ & $e^-\mu^+\to \nu_e\bar{\nu}_\mu W^-(\to \mu^-\bar\nu_\mu) W^+(\to e^+\nu_e) Z/\gamma^*(\to\mu^+\mu^-)$  & $2.3\times10^{-7}$ \\
        \hline\hline
    \end{tabular}}
    \caption{Same as in Table~\ref{tab:alp_hadtaus} for signal selection involving leptonic $\tau$ decays. $\tau_\ell$ denotes \mbox{$\tau\to\ell \nu\bar\nu$} ($\ell = e,\mu$).}
    \label{tab:alp_leptaus}
\end{table}

As already mentioned, modern detectors are expected to yield a large $\tau$-jet tagging efficiency. For definiteness, we set such a parameter in \textsc{Delphes} to be 90\% for taus with $p_T > 10$~GeV. In fact, the identification efficiency is expected to degrade substantially in case of soft jets. Furthermore, the geometric acceptance of the detector can severely reduce the selection efficiency of the signal, especially in asymmetric $e^-\mu^+$ collisions, as discussed above in the case of searches for $e^-\mu^+ \to a\gamma$. As shown in the right plot of Figure~\ref{fig:cs_taustate_ma}, the combination of these effects limit the signal efficiency to $\varepsilon_\text{S} \lesssim 10\%\,(25\%)$ for $e^-\mu^+$ ($\mu^+\mu^+$) collisions. In addition, no sensitivity is expected for ALP masses smaller than $10-20$~GeV, because this results in hadronic taus that are too soft to be detected, as mentioned above.

Due to the limited sensitivity to light ALPs, we also consider signal selection based on leptonic $\tau$ decays, for which $\sigma_\text{S}$ is reduced by a factor $\left[\text{BR}(\tau\to\ell\nu\bar\nu)\right]^2 \approx 3\%$ (times the branching fraction of the selected $a\to \ell\tau$ mode) relative to the production cross section displayed in the left panel of Figure~\ref{fig:cs_taustate_ma}.

The cross sections of the BG processes relevant to ALP searches involving hadronic and leptonic taus are displayed, respectively, in Table~\ref{tab:alp_hadtaus} and Table~\ref{tab:alp_leptaus}. As we can see, searches for $\tau$ LFV ALPs at $\mu$TRISTAN operating as a $e^-\mu^+$ collider are expected to be background-free for $\mathcal{L} =10~\text{fb}^{-1}$. In contrast, the SM background for the searches based on $\mu^+\mu^+$ is non-negligible, but only gives $\mathcal{O}(10)$ events. Following from Eq.~\eqref{eqn:significance}, this mildly limits the resulting sensitivity on the $\mu$-$\tau$ ALP couplings.

\begin{figure}[t!]
    \centering
    \subfigure[LFV ($\mu$-$\tau$) couplings only]{\includegraphics[width=0.48\textwidth]{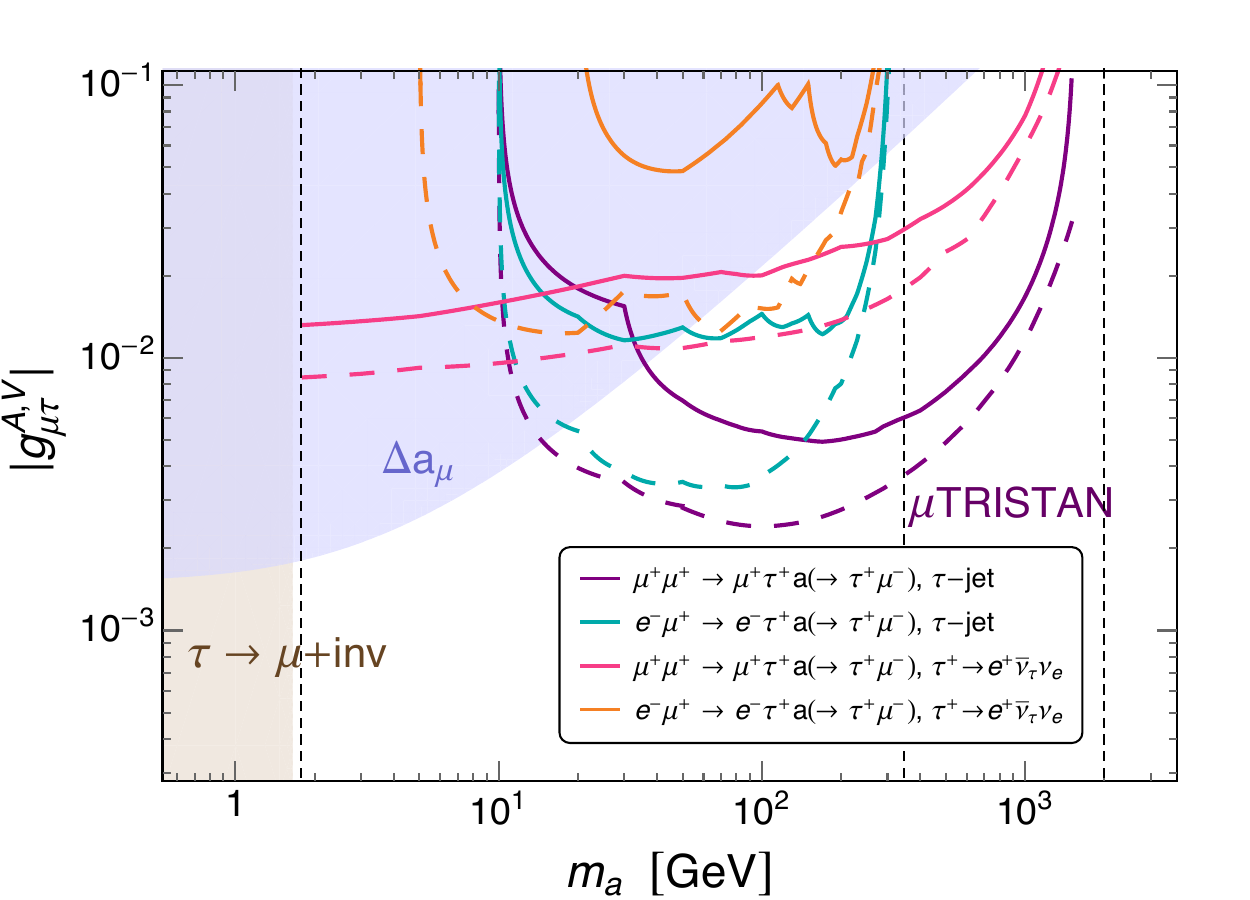}}
    \hfill
    \subfigure[LFV ($\mu$-$\tau$) and LFC couplings]{\includegraphics[width=0.48\textwidth]{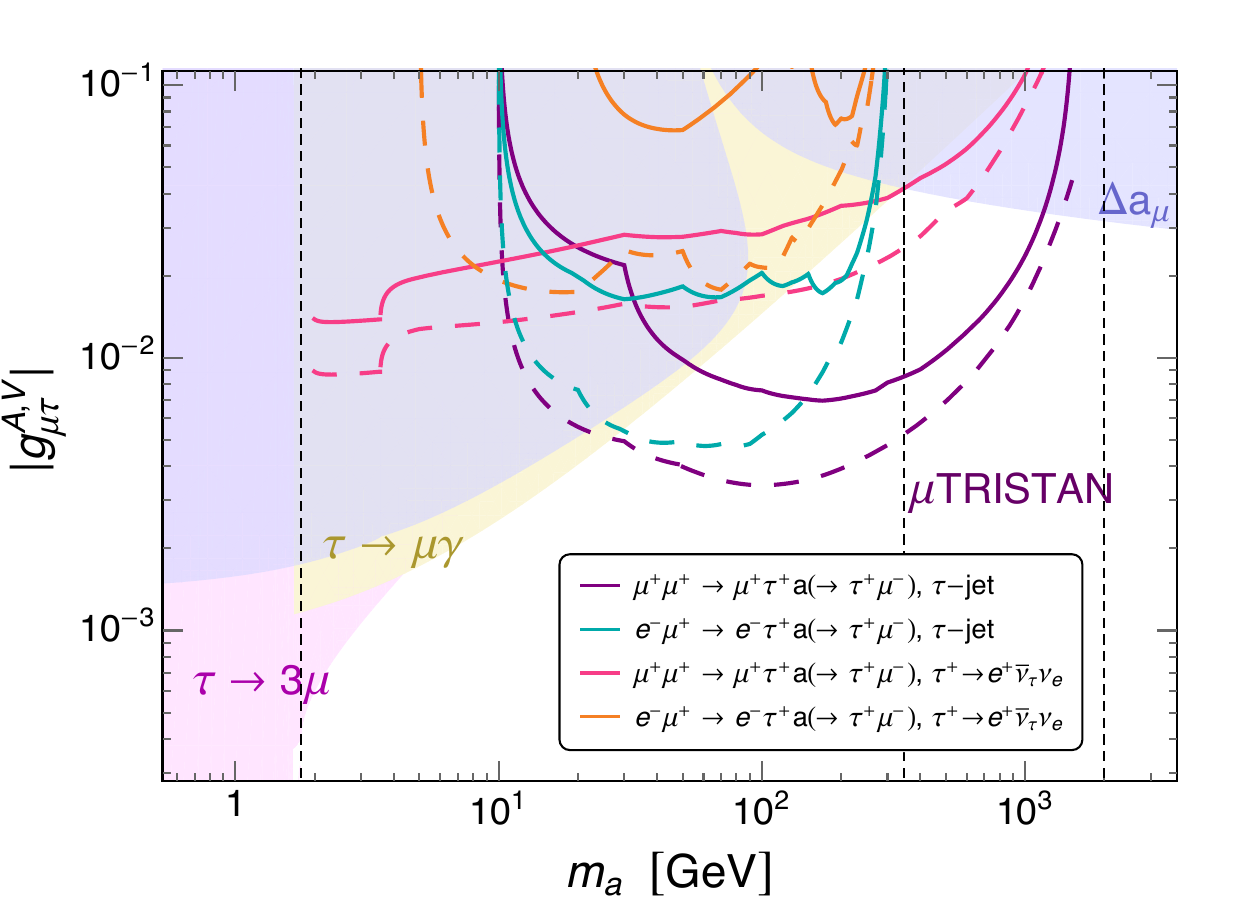}}
    \subfigure[LFV ($e$-$\tau$) couplings only]{\includegraphics[width=0.48\textwidth]{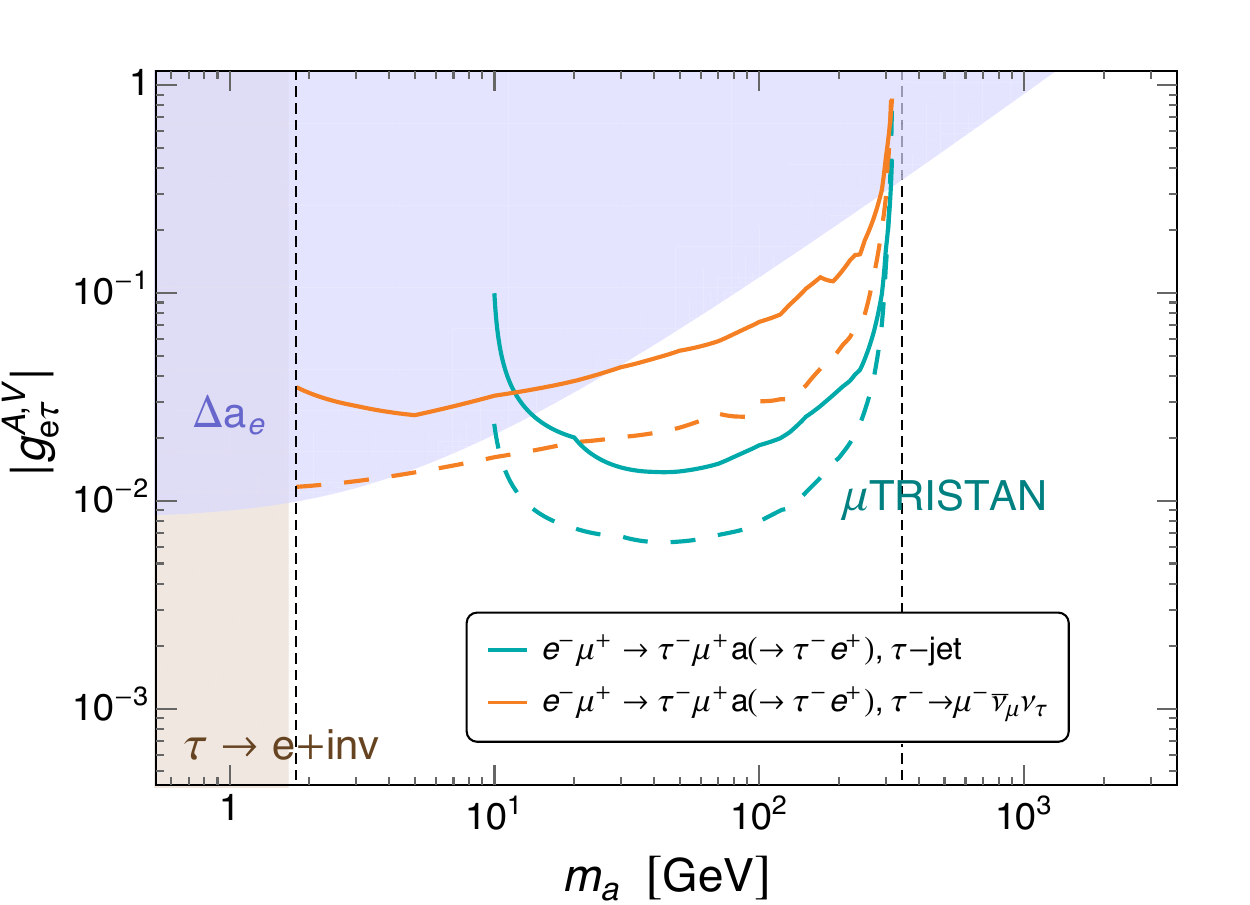}
    }
    \hfill
    \subfigure[LFV ($e$-$\tau$) and LFC couplings]{\includegraphics[width=0.48\textwidth]{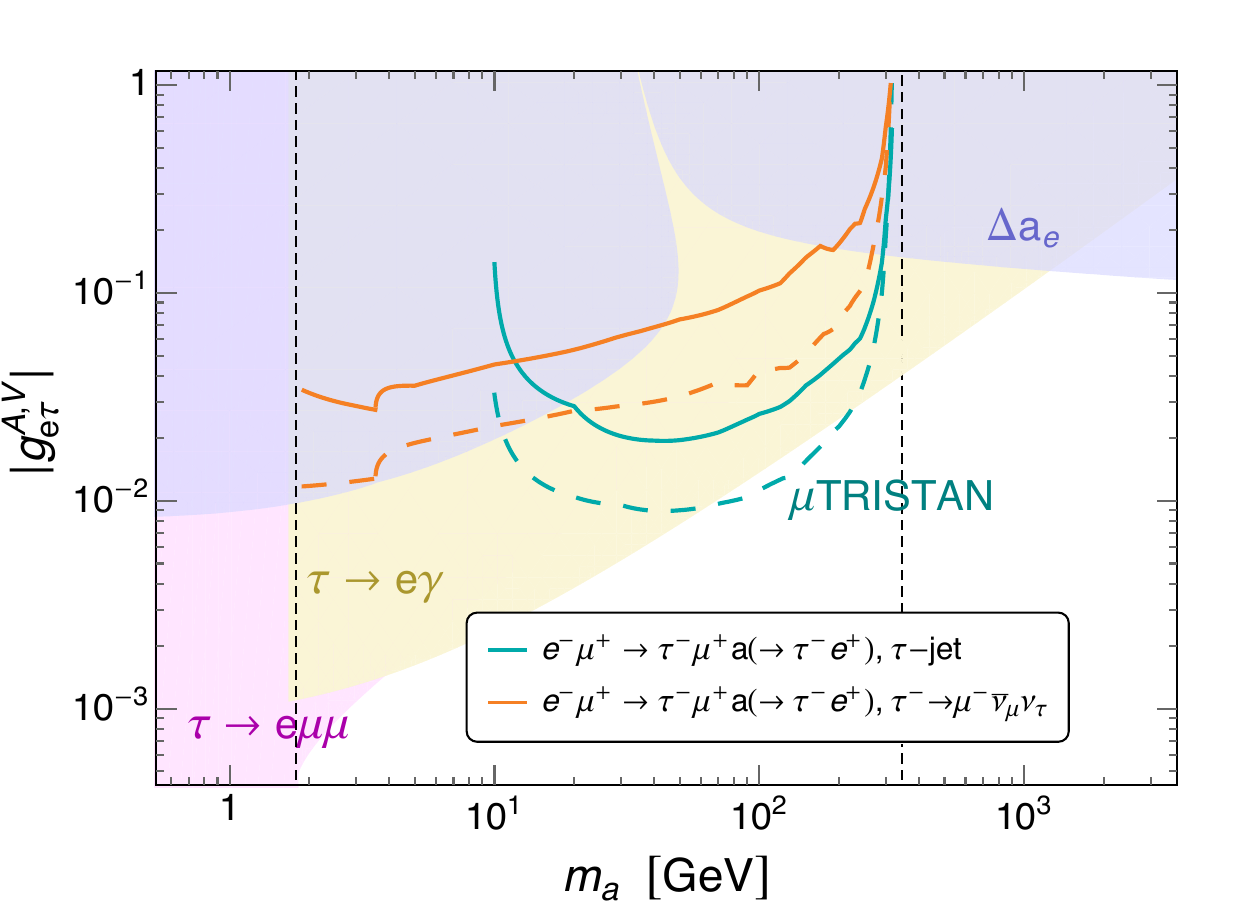}
    }
    \caption{Sensitivity of $\mu$TRISTAN to ALP production through LFV interactions with $\tau$ leptons. In the first row, the expected 95\% CL exclusion limits on $|g^{V,A}_{\mu\tau}|$ as a function of $m_a$ are shown for a purely LFV ALP with $C^A_{\mu\tau}=C^V_{\mu\tau}$ (left panel) and an ALP with LFC and LFV couplings and $C^A_{\ell\ell}=C^{A,V}_{\mu\tau}$ (right panel). Solid (dashed) lines denote sensitivities of a detector with $|\eta|_{\text{max}} = 2.5$ ($|\eta|_{\text{max}} = 3.5$).
    Analogously, the plots in the second row show limits on $|g^{V,A}_{e\tau}|$ for the case $C^A_{e\tau}=C^V_{e\tau}$ (left panel) and $C^A_{\ell\ell}=C^{A,V}_{e\tau}$ (right panel).
    The coloured regions are excluded by low-energy leptonic processes: two-body tau decay into muon or electron and an invisible boson (brown), $\tau \to \ell\ell\ell^{(\prime)}$ decays (pink), $\tau\to e\gamma$ (yellow), the ALP contribution to muon or electron $g-2$, respectively, $\Delta a_\mu$ and $\Delta a_e$ (purple). See main text for details.}
    \label{fig:gelltau}
\end{figure}

\begin{figure}[t!]
    \centering
    \subfigure[LFV ($\mu$-$\tau$) and LFC couplings]{\includegraphics[width=0.48\textwidth]{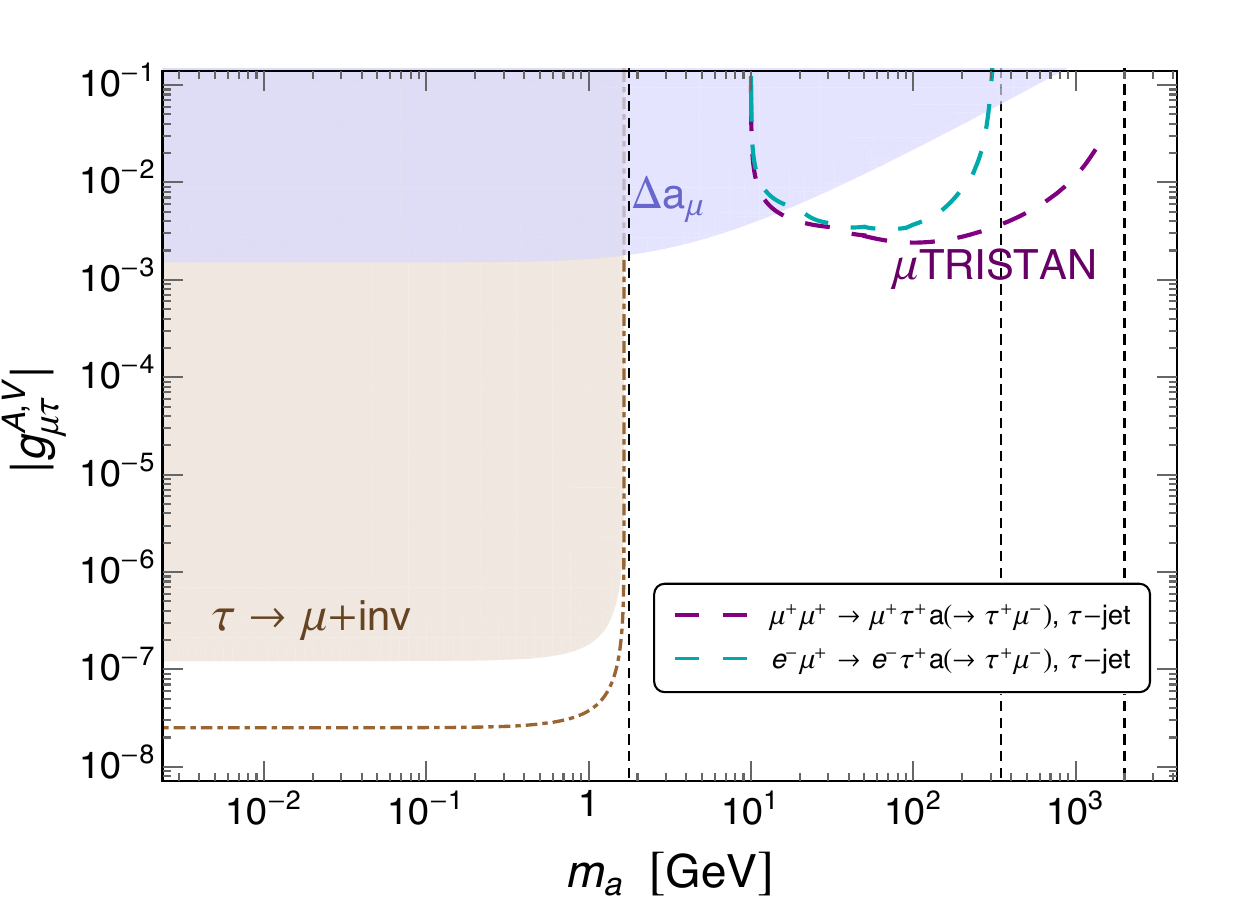}}
    \hfill
    \subfigure[LFV ($\mu$-$\tau$) and LFC couplings]{\includegraphics[width=0.48\textwidth]{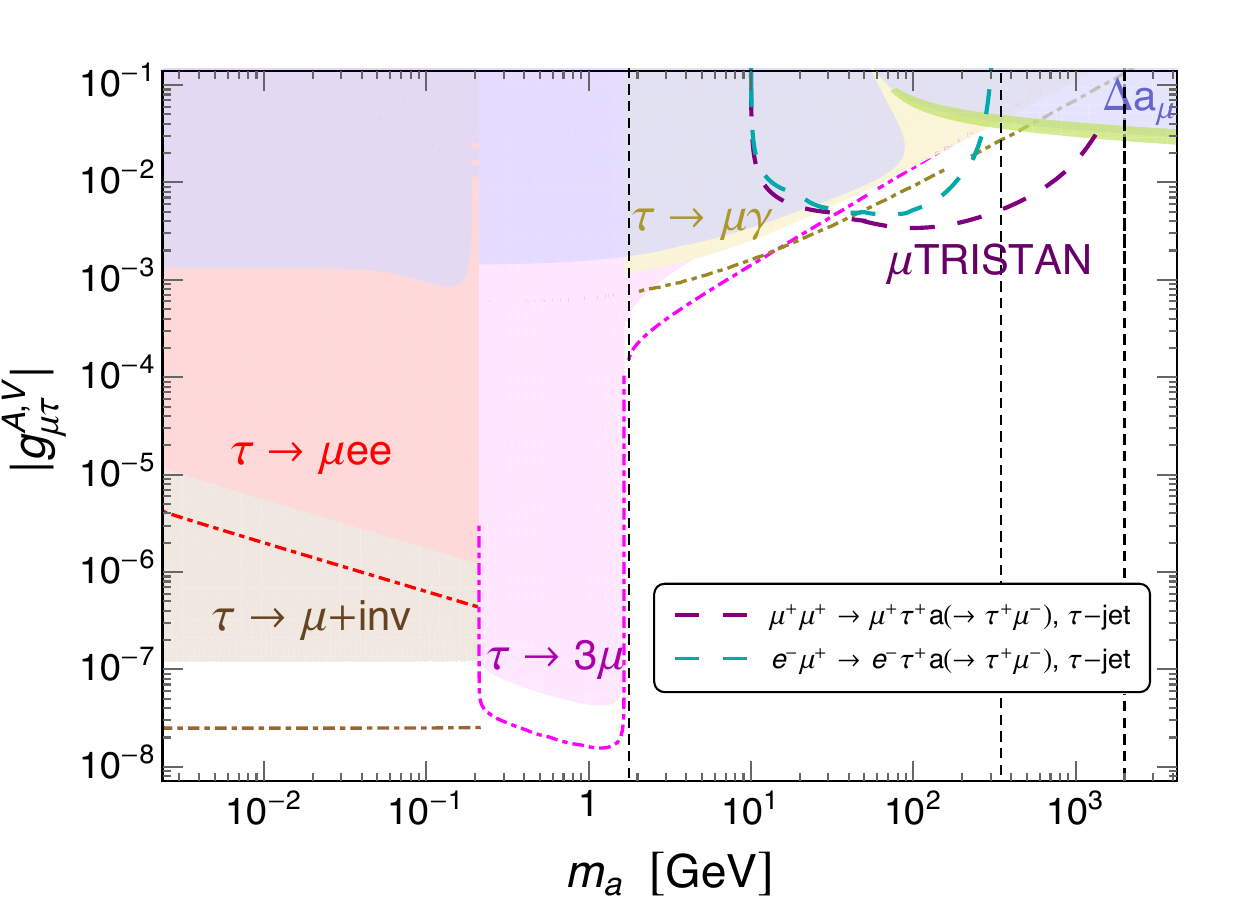}}
        \subfigure[LFV ($e$-$\tau$) and LFC couplings]{\includegraphics[width=0.48\textwidth]{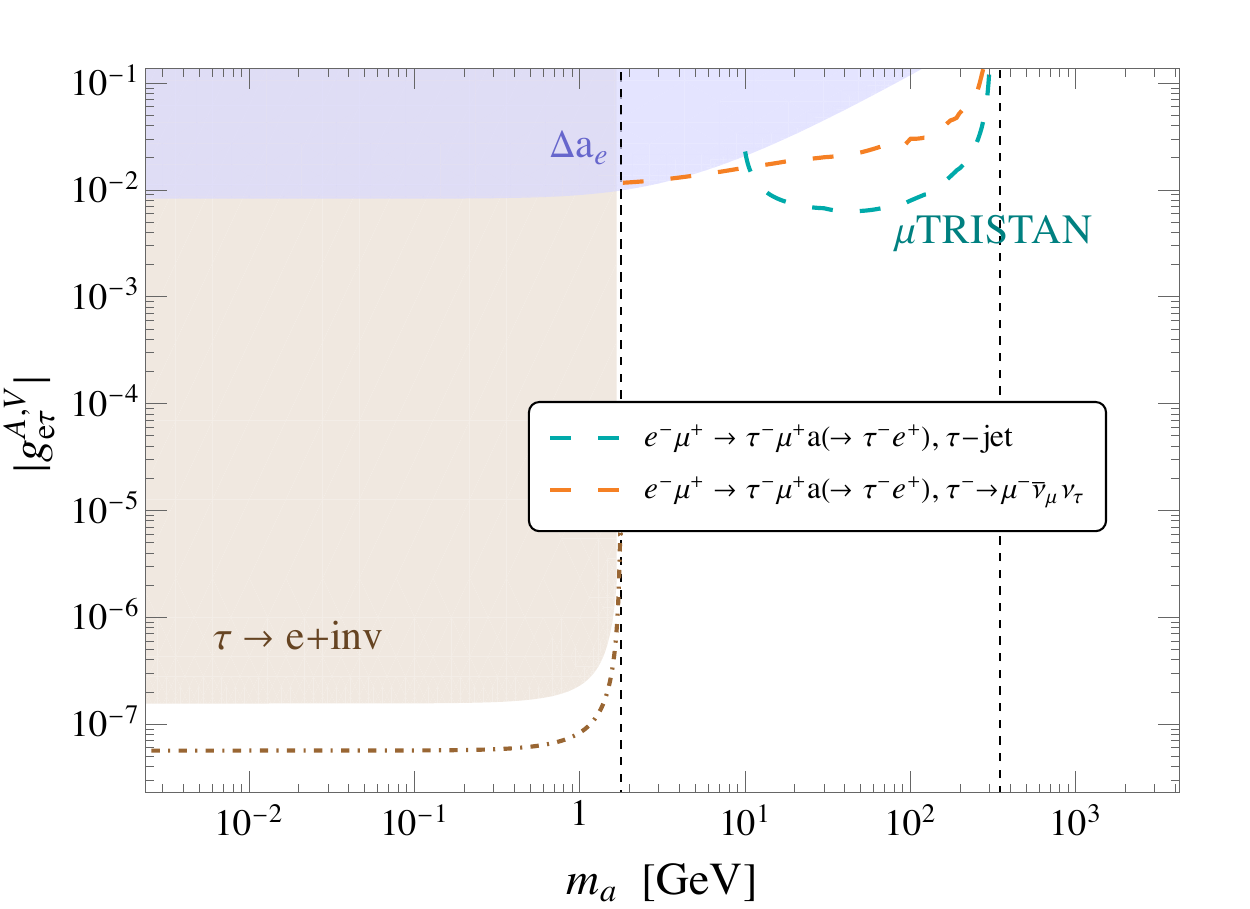}}
    \hfill
    \subfigure[LFV ($e$-$\tau$) and LFC couplings]{\includegraphics[width=0.48\textwidth]{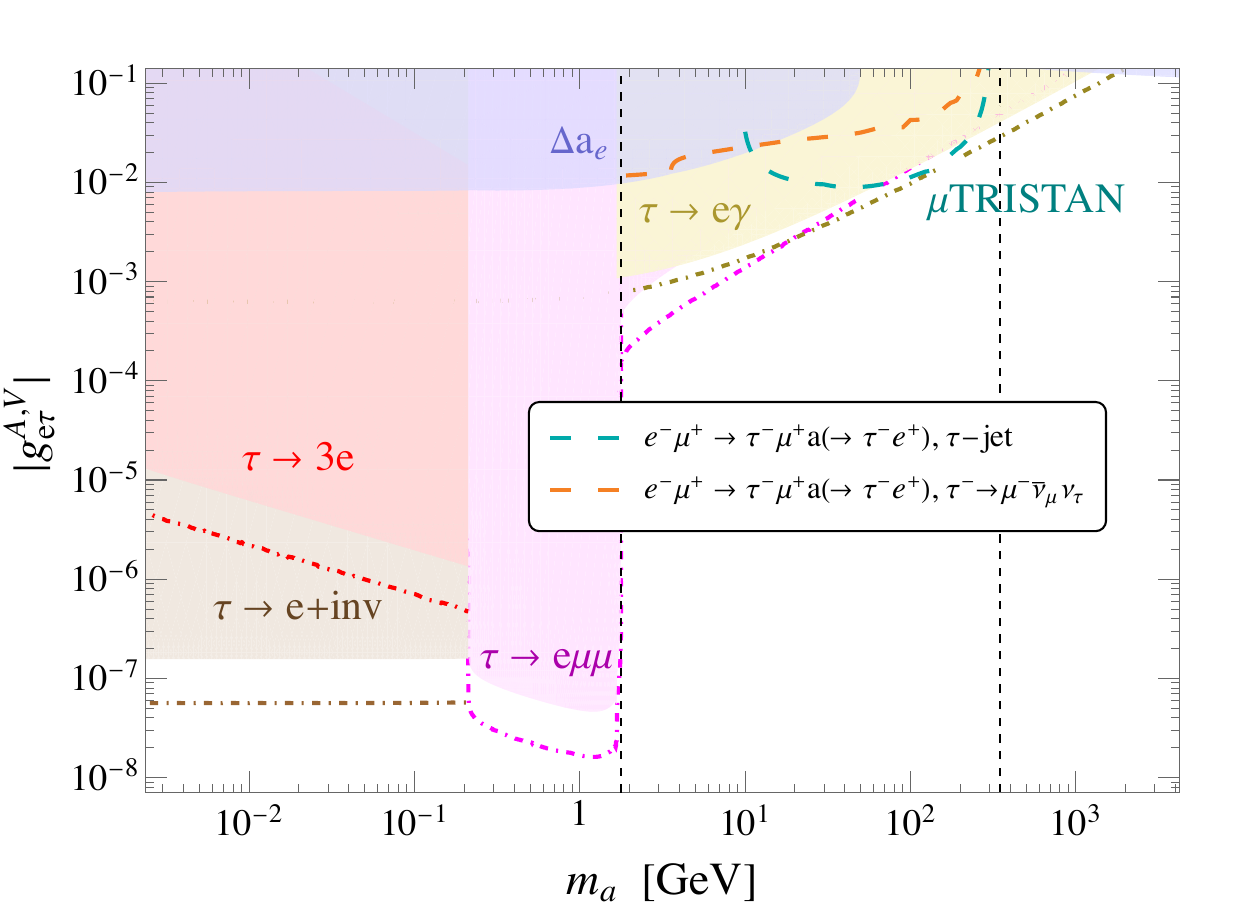}}
    \caption{Overview of the present and future constraints on LFV ALPs in the $m_a-|g^{V,A}_{\mu\tau}|$ plane (first row) and $m_a-|g^{V,A}_{e\tau}|$ plane (second row) compared with the $\mu$TRISTAN sensitivity as estimated in this work. The model's parameters and the colour coding are as in Figure~\ref{fig:gelltau}. The dot-dashed lines represent the expected Belle~II sensitivities on LFV $\tau$ decays with $\mathcal{L}=50~\text{ab}^{-1}$. See main text for details.}
    \label{fig:constraint_gelltau}
\end{figure}

\paragraph{$\mu$TRISTAN sensitivity.} 
 The expected sensitivity of the $\mu$TRISTAN searches discussed in this section is summarised in Figure~\ref{fig:gelltau}. As before, the solid (dashed) lines denote the lower limits on the relevant LFV couplings that can be obtained by a detector with geometric acceptance $|\eta|_\text{max} =2.5$ ($|\eta|_\text{max} =3.5$) after collecting $\mathcal{L}=10~\text{ab}^{-1}$ of data. 
 Searches based on leptonic $\tau$ decays can provide some sensitivity to the LFV couplings for $m_a \lesssim 10$~GeV. For heavier ALPs, however, searches for signatures with $\tau$ jets have a better sensitivity and can constrain couplings at the $10^{-3}-10^{-2}$ level, especially for $m_a = \mathcal{O}(100)$~GeV, as expected from the signal efficiencies plotted in Figure~\ref{fig:cs_taustate_ma} (right).
Comparing the $\mu$TRISTAN sensitivity with the bounds from low-energy observables (LFV decays and leptonic $g-2$), one can see that $\mu$TRISTAN can test portions of the parameter space that are currently unconstrained, especially if the ALP exclusively (or dominantly) couples with leptons of different flavours. Our results show that, in addition, some sensitivity to $\mu$-$\tau$ interactions beyond the present low-energy limits is possible even if LFC couplings are present. Nevertheless, to this end, it is important to design a detector with a geometric acceptance of light leptons and hadronic taus up to $|\eta|_\text{max} = 3.5$ or higher.
In the $e$-$\tau$ case, our results show that only a marginal improvement beyond the bounds on LFV $\tau$ decays can be expected. 
In the following, we briefly discuss the limits set by LFV decays and muon/electron $g-2$. For further information, we refer to Appendix~\ref{app:LFV} where the expressions used to compute them are collected together with the current and forecast experimental limits.

\paragraph{Present and future LFV and $g-2$ constraints.} 
As shown in Figure~\ref{fig:gelltau}, if the ALP does not couple to same-flavour leptons and is too heavy for the $\tau \to \ell\,a$ decays to be allowed, the only low-energy constraints that affect our parameter space stem from the anomalous magnetic moments of the leptons. 
Indeed, in the presence of $\mu$-$\tau$ ($e$-$\tau$) ALP interactions, an ALP-tau loop provides a non-standard contribution $\Delta a_\mu$ ($\Delta a_e$) to the muon (electron) $g-2$. Although the theoretical predictions within the SM are currently subject to debate for either observables, a model giving too large values of $|\Delta a_\mu|$ or $|\Delta a_e|$ is certainly excluded. 
Here, we conservatively show the $2\sigma$ exclusion (denoted as a purple region) obtained adopting in each case the SM prediction that is in better agreement with the experimental measurement. These are, for the muon $g-2$, the prediction based on the most precise determination of the hadronic vacuum polarization~(HVP) from lattice QCD~\cite{Borsanyi:2020mff} and, for the electron $g-2$, the prediction based on the value of the fine-structure constant measured by means of matter-wave interferometry of rubidium atoms ~\cite{Morel:2020dww}\,---\,see Appendix~\ref{app:g-2} for details. In Figure~\ref{fig:constraint_gelltau}, we also show as a green band the values of the parameters that could explain (at the $2\sigma$ level) the recent measurement of the Muon g-2 Experiment~\cite{Muong-2:2023cdq} if instead the data-driven approach to the HVP using dispersion relations is employed~\cite{Aoyama:2020ynm}. As we can see, that region is partly within the reach of our estimated $\mu$TRISTAN sensitivity.

If $\tau$ decays into ALPs are kinematically open and/or in the presence of LFC couplings, a number of LFV $\tau$ decays set important constraints on our parameter space. This is better depicted in Figure~\ref{fig:constraint_gelltau}, where the expected future limits from the Belle~II experiment~\cite{Belle-II:2018jsg,Banerjee:2022vdd} are also shown as dot-dashed lines.
In particular, we show as brown regions the current best limits (in the $10^{-4}-10^{-3}$ range) obtained by Belle~II on the branching ratios of $\tau\to \ell\,a$, with an invisible $a$~\cite{Belle-II:2022heu}, alongside with the future sensitivity \mbox{$\text{BR}(\tau\to \ell\,a)\simeq 10^{-5}$} that can be achieved by the same experiment with the expected full data set of $50~\text{ab}^{-1}$~\cite{Calibbi:2020jvd}. 
As we can see from the figures, LFC couplings of the ALP to electrons and muons can also mediate a number of LFV decays of the kind $\tau\to \ell\ell\ell^{(\prime)}$ ($\ell,\ell^\prime = e,\mu$) that, in the light ALP regime can be more or less constraining than the searches for an invisible ALP depending on the lifetime of the latter. Interestingly, the future Belle~II limits $\text{BR}(\tau \to \mu\mu\mu) < 3.6\times 10^{-10}$ and
$\text{BR}(\tau \to e\mu\mu) < 4.5\times 10^{-10}$~\cite{Banerjee:2022vdd} can set more stringent bounds than, respectively, $\tau \to \mu\gamma$ and $\tau \to e\gamma$ also if the ALP is off-shell, as long as $m_a \lesssim 20$~GeV, and show a nice complementarity with the sensitivity that can be achieved at $\mu$TRISTAN.

\section{Summary and conclusions}
\label{sec:concl}

Axions and ALPs are intriguing candidates for NP beyond the SM. If these particles couple to the lepton sector of the SM, there is no fundamental reason why their interactions should be flavour conserving. In fact, LFV ALP interactions are predicted within a wide range of explicit UV models~\cite{Calibbi:2020jvd}. The recently proposed $\mu$TRISTAN $e^-\mu^+$ (and $\mu^+\mu^+$) collider~\cite{Hamada:2022mua}, drawing from the mature $\mu^+$ beam technology developed for the J-PARC muon $g-2$ experiment~\cite{Abe:2019thb}, plays as an ideal machine to explore possible NP in the charged lepton sector. In this work, we studied the search potential for LFV ALPs at $\mu$TRISTAN with different beams and energies. We assessed the sensitivity to the LFV ALP couplings that can be achieved employing a number of ALP production channels and compared it with the existing low-energy leptonic constraints as well as future improvements. Our main results can be summarised as follows.
\begin{itemize}
\item The most promising ALP production modes at $\mu$TRISTAN include $e^- \mu^+ \to a \gamma$, $e^- \mu^+ \to e^- \tau^+ a$, $\mu^+ \mu^+ \to \mu^+ \tau^+ a$, and $e^- \mu^+ \to \tau^- \mu^+ a$. They are induced by the LFV ALP couplings $C_{e\mu}^{A,V}$, $C_{\mu\tau}^{A,V}$, and $C_{e\tau}^{A,V}$, respectively. We examined the $\mu$TRISTAN sensitivity to these couplings by means of a cut-based collider analysis.
\item For each signal channel, the analysis is divided into two mass regimes for the ALP and limits are set on dimensionless coupling $|g_{ij}^{V,A}| \simeq m_j C_{ij}^{V,A}/2f_a$ at an integrated luminosity of $\mathcal{L} = 10$ ab$^{-1}$. The limits scale as $\sim\left(10~\text{ab}^{-1}/\mathcal{L}\right)^{1/2}$ for background free signals (that are usually the most constraining ones). 
\item We find that for improved signal sensitivity, a wider geometric acceptance of the detector is desirable in the case of the asymmetric $e^-\mu^+$ collisions, as the ALP decay products tend to be collinear to the muon beam direction with large pseudorapidity. In our analysis, we compare limits on the dimensionless couplings for $|\eta| < 2.5$ and $|\eta| < 3.5$ respectively.
\item \textbf{Light ALP.} When $m_a<m_\ell + m_{\ell'}$ with $m_\ell<m_{\ell'}$, the ALP is either long-lived or may promptly decay to $\ell^+\ell^-$ depending if the LFC coupling $C^A_{\ell\ell}$ is allowed. For the $e^- \mu^+ \to a \gamma$ channel, we find that the parameter space that $\mu$TRISTAN can probe has already been excluded by low-energy LFV processes such as \text{Mu}--$\overline{\text{Mu}}$ oscillation, $\mu\to e+{\rm inv.}$ and $\mu\to eee$. For the ALP productions with $\tau$ lepton, we are not able to place a limit on the LFV $\tau$ couplings because of the non-negligible SM background. However, analogous LFV $\tau$ decays already strongly constrain this regime.
\item \textbf{Heavy ALP.} In the heavier ALP regime, i.e.~when $m_a>m_\ell + m_{\ell'}$, under different assumptions of leptonic couplings, the ALP can decay to either flavour changing or flavour conserving lepton final states. 
\begin{itemize}
\item The search for heavy ALP through $e^- \mu^+ \to a \gamma$ can test the parameter space beyond the limits from \text{Mu}--$\overline{\text{Mu}}$ oscillation and $\mu\to e \gamma$ and has the potential to test the $|g^{V,A}_{e\mu}|$ couplings to values as low as $\sim \mathcal{O}(10^{-4})$.
\item For the associated productions of ALP--$\tau$--$\ell$ with $m_a>m_\tau + m_\ell$, the leptonic (hadronic) $\tau$ decay mode is sensitive to relatively low (high) ALP mass values. Signatures with hadronic $\tau$ decays are able to probe values of the $g_{\ell\tau}^{V,A}$ couplings currently allowed by $\tau$ LFV decay and leptonic $g-2$ and reach a sensitivity of $|g_{\mu\tau}^{V,A}| \sim 10^{-3}-10^{-2}$ and $|g^{V,A} _{e\tau}| \sim 10^{-2}$ for $m_a = \mathcal{O}(100)$~GeV.
\end{itemize}
\end{itemize}

The $\mu$TRISTAN sensitivity reach summarised above approximately corresponds to the following limits on the cut-off scale of the relevant operators in Eq.~(\ref{eq:L_ALP1}):
\begin{align}
{f_a}/{C^{V,A}_{e\mu}} \lesssim 500~\text{GeV}\,,\quad
{f_a}/{C^{V,A}_{\mu\tau}} \lesssim 300~\text{GeV}\,,\quad
{f_a}/{C^{V,A}_{e\tau}} \lesssim 150~\text{GeV}\,.
\end{align}
Comparing these results with the $\mu$TRISTAN $e^-\mu^+$ mode centre of mass energy, $\sqrt{s} \simeq 346$~GeV, we can conclude that our model-independent effective field theory approach is marginally justified in the case of searches for $e$-$\mu$ violation. On the other hand, the $\mu$TRISTAN  sensitivity on $\mu$-$\tau$ and, especially, $e$-$\tau$ ALP LFV searches would be within the validity of the effective field theory only if $C^{V,A}_{\ell\tau} >1$, which would point to an ALP originating from a strongly-coupled sector.
The  search for  $\mu$-$\tau$  LFV with $\sqrt{s} = 2$~TeV $\mu^+\mu^+$ collisions is definitely not sensitive to ranges of couplings for which our effective field theory is consistent.

It is a model-dependent question whether the fields associated with the UV completion substantially affect the ALP phenomenology. Likewise, LHC limits on the mass of new particles charged under the SM electroweak interactions depend on the specific production and decay modes and range approximately between $100$~GeV and $1$~TeV, hence it will depend on the UV completion whether $\mu$TRISTAN is capable to test ALP models beyond the LHC constraints or not. While this important discussion is beyond the scope of our work, we just notice here that a UV-complete ALP model might entail fields charged under the SM gauge group at a scale parametrically larger than $f_a$, which is the vev of the singlet scalar field that breaks the associated $\text{U}(1)$ symmetry.\footnote{This is for instance the case of the ``familon'', the PNGB arising from a Froggatt-Nielsen $\text{U}(1)$ flavour symmetry. The UV completion of such flavour models (e.g.~consisting in vectorlike fermions) typically lie at a scale $\Lambda/f_a \gtrsim \mathcal{O}(10)$, see e.g.~\cite{Calibbi:2020jvd}.}

\begin{figure}[t!]
\centering
\subfigure[LFV ($e$-$\mu$) couplings only]{
\includegraphics[width=0.48\textwidth]{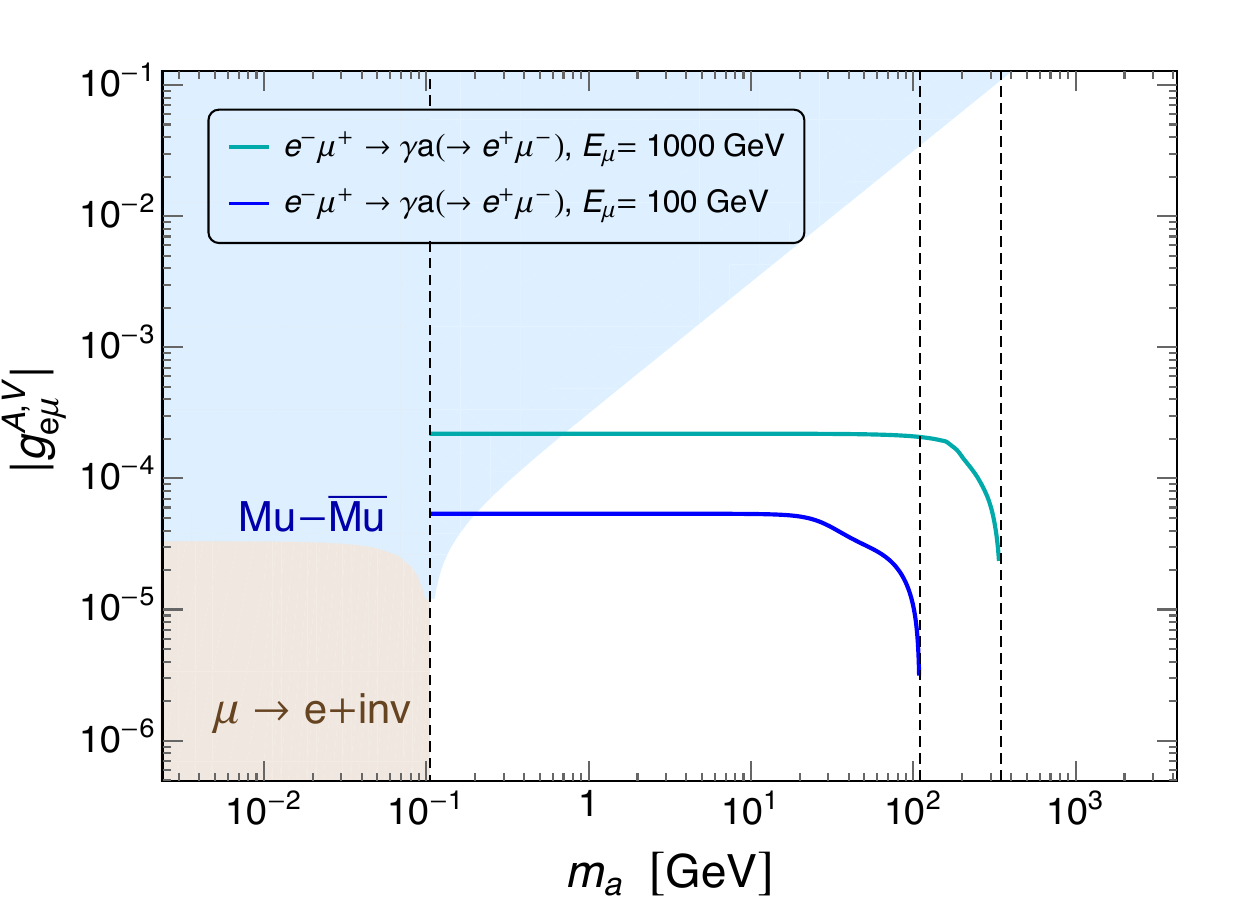}}
\hfill
\subfigure[LFV ($e$-$\mu$) and LFC couplings]{
\includegraphics[width=0.48\textwidth]{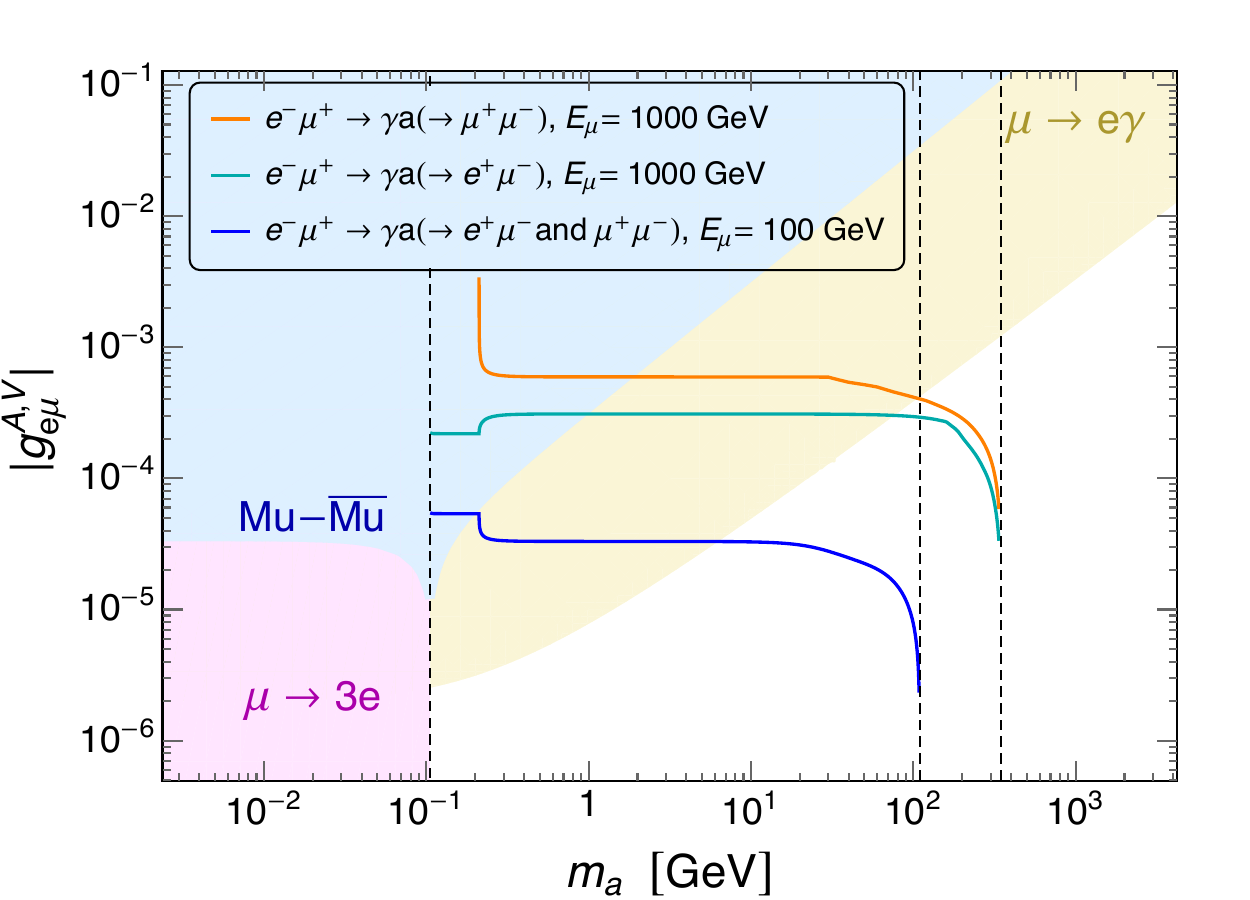}}
\caption{Comparison of the limits on the ALP couplings to $e$-$\mu$ that could be achieved at $\mu$TRISTAN with different $\mu^+$ beam energy configurations, while keeping $E_e = 30$~GeV: $E_\mu=1000$~GeV ($\sqrt{s} \simeq 346$~GeV) and $E_\mu=100$~GeV ($\sqrt{s} \simeq 110$~GeV). 
Note that the background from $e^-\mu^+\to \gamma\mu^+\mu^-+$~MET is negligible for $E_\mu=100$~GeV  ($\sigma_\text{B} =4.8\times10^{-4}$~fb), so that we can combine the signal from the LFV ALP decay with the LFC one. The model's parameters and colour coding are as in Figure~\ref{fig:constraint_gemu_zoom_to_collider}.}
\label{fig:different_muon_beam}
\end{figure}

As a future outlook, one can envisage different muon beam energy configurations for the $\mu$TRISTAN project. We have seen that the ALP production cross-section for the $e^-\mu^+$ collisions increases for lower c.m.~energies, cf.~Eq.~\eqref{eq:cross-sec-emu-alpgamma}. Furthermore,
if the asymmetry in the beam energies is lowered, the detecting efficiency of the signal also increases, as the ALP decay products are less collinear to the muon beam. We find that the optimal choice for the beam energy configuration is $(E_e, E_\mu) = (30~\rm{GeV},100~\rm{GeV})$,\footnote{Such a configuration corresponds to $\sqrt{s}\simeq 110$~GeV, hence it is clearly beyond the main purpose of $\mu$TRISTAN to work as a Higgs factory. However, if evidence of light NP in the $\sqrt{s} \simeq 346$~GeV run is observed, performing precision studies of it at lower $\sqrt{s}$ is a reasonable option.} and study how this influences the collider bound on $g_{e\mu}^{V,A}$ that can be obtained in the search for $e^-\mu^+\to \gamma a(\to e^+\mu^-/ \mu^+\mu^-)$. Although in such a case the luminosity is expected to be lower due to relativistic effects\,---\,for more details see~\cite{Hamada:2022mua}\,---\,we show in Figure~\ref{fig:different_muon_beam} the improvement on the sensitivity to $g_{e\mu}^{V,A}$ (by approximately one order of magnitude) assuming that an integrated luminosity of $10~\rm{ab}^{-1}$ can still be reached in the future with advanced beam intensity and focusing technologies. 
We notice that this improved sensitivity corresponds to ${f_a}/{C^{V,A}_{e\mu}}$ above the TeV scale. In combination with a lower value of $\sqrt{s}$, this enlarges the validity range of the the effective Lagrangian in Eq.~(\ref{eq:L_ALP1}).
For the ALP production modes with an associated $\tau$ lepton in the final state, we observe that $\mu$TRISTAN with the above energy configuration does not exhibit a better performance compared to the results obtained in this paper, although the cross sections also increase, since the $\tau$-jet identification efficiency rapidly deteriorates for softer jets. 

\subsection*{Acknowledgements}
We would like to thank organisers and participants\,---\,especially, Chih-Ting Lu, Zuowei Liu, and Jin Sun\,---\,of the 2nd  TOPAC meeting (Southeastern University, Nanjing, May 30th - June 2nd 2024), where LM presented our preliminary results, for useful feedback and valuable discussions. We also thank Jin Sun for pointing out a typo and a mistaken sign in a previous version of the manuscript.
LC and TL acknowledge financial support from the National Natural Science Foundation of China (NSFC) under the grant No.~12035008. LC is additionally supported by the NSFC under the grant No.~12211530479.
TL is also supported by the NSFC (Grants No.~12375096, 11975129) and ``the Fundamental Research Funds for the Central Universities'', Nankai University (Grant No.~63196013).


\appendix 

\section{ALP decay into photons}
\label{app:gamma}

The $a\to\gamma\gamma$ decay width can be written as~\cite{Bauer:2017ris,Bauer:2018uxu}
\begin{align}
    \Gamma(a\to\gamma\gamma)=\frac{m_a^3 \alpha ^2 }{64 \pi ^3 f_a^2}\left[E_{\text{UV}}+\sum_{i}C_{\ell_i\ell_i}^A B\left( \frac{4m_{\ell_i}^2}{m_a^2}\right)\right]^2,
\end{align}
where the loop function $B$ reads
\begin{align}
    B(x)= 1-x f(x)^2,~~ f(x)=\begin{cases}
 \arcsin\left(\frac{1}{\sqrt{x }}\right) & x\geq 1 \,,\\
 \frac{\pi }{2}+\frac{i}{2} \log \left(\frac{\sqrt{1-x }+1}{\sqrt{1-x}+1}\right) & x <1 \,.
\end{cases}
\label{eq:B}
\end{align}
In the above expression, $E_{\text{UV}}$ is a model-dependent UV ALP-photon coupling (related to the electromagnetic anomaly of the global $\text{U}(1)$ symmetry) and the second term is due to lepton loops that are unavoidable if LFC couplings exist. Unless otherwise noted, we only consider the latter model-independent contribution, that is, we set $E_{\text{UV}}=0$ throughout the paper.

\section{Low-energy leptonic processes}
\label{app:LFV}

\subsection{Muonium antimuonium oscillations}

The ALP mediated $\text{Mu}-\overline{\text{Mu}}$ transition probability is given by~\cite{Endo:2020mev,Bauer:2021mvw} 
\begin{align}
    P_{\text{Mu}\,\overline{\text{Mu}}}=\frac{m_\mu^4}{2\pi^2 a_B^6 \Gamma_\mu^2 [(m_\mu^2-m_a^2)^2+\Gamma_a^2 m_a^2] f_a^4} & \left[|c_{0,0}|^2\left|(C^V_{e\mu})^2-\left(1+\frac{1}{\sqrt{1+X^2}}\right)(C^A_{e\mu})^2\right|^2 + \right.\nonumber\\
    &\left. |c_{1,0}|^2\left|(C^V_{e\mu})^2-\left(1-\frac{1}{\sqrt{1+X^2}}\right)(C^A_{e\mu})^2\right|^2\right],
    \label{eq:Pmumu}
\end{align}
where $\Gamma_\mu \simeq 3.00 \times 10^{-19}$~GeV is the muon decay rate, and $a_B \simeq 2.69 \times 10^5~\text{GeV}^{-1}$ is the muonium Bohr radius, while $X$ is a parameter related to the magnetic field $B$ employed in the experimental apparatus, defined as $X=6.31~(B/1\,\text{T})$. The field also affects the probability of populating the initial state with angular momentum $(J,m_{J})$, which is expressed by the quantity $|c_{J,m_{J}}|^2$. Compared to~\cite{Endo:2020mev}, we include here the complete ALP propagator to ensure the validity of the formula for both light and heavy ALPs.

The most precise search for muonium oscillations to date was performed by the MACS experiment, which used $B=0.1$~T, for which $|c_{0,0}|^2=0.32$ and $|c_{1,0}|^2=0.18$. The 90\%~CL upper limit on the oscillation probability obtained by MACS is $P_{\text{Mu}\,\overline{\text{Mu}}}<8.3\times 10^{-11}$~\cite{Willmann:1998gd}. Concerning the future prospects, the proposed MACE experiment is expected to reach  with \mbox{$5\times 10^8\,\mu/s$} on target and one year of data taking) a single event sensitivity of $3\times 10^{-14}$, corresponding to the 90\%~CL upper limit \mbox{$P_{\text{Mu}\,\overline{\text{Mu}}}<7\times 10^{-14}$}~\cite{Bai:2024skk,Bai:2022sxq,mace_talk}. MACE plans to employ the same field as MACS, $B=0.1$~T, hence the quantities $X$ and $c_{J,m_{J}}$, required in Eq.~\eqref{eq:Pmumu} to interpret the experimental results in terms of the ALP parameters, are also the same.

\subsection{LFV muon and tau decays}

\begin{table}[h!]
\renewcommand{\arraystretch}{1.4}    
\centering
    \begin{tabular}{ccc}
    \hline\hline
    Process & Current limit & Future limit \\ \hline\hline
    $\mu\to e\gamma$ & $4.2\times 10^{-13}$~MEG~\cite{MEG:2016leq} & $6\times 10^{-14}$~MEG~II~\cite{MEGII:2018kmf} \\
       $\mu\to ee\bar e$  & $1.0\times 10^{-12}$~SINDRUM~\cite{Bellgardt:1987du} & $10^{-16}$~Mu3e~\cite{Blondel:2013ia}\\ \hline
       $\tau\to \mu\gamma$ & $4.2\times 10^{-8}$~Belle~\cite{Belle:2021ysv} & $6.9\times 10^{-9}$~Belle~II~\cite{Belle-II:2018jsg,Banerjee:2022vdd}\\
       $\tau\to \mu\mu\bar\mu$ & $1.9\times 10^{-8}$~Belle~II~\cite{Belle-II:2024sce}  & $3.6 \times 10^{-10}$~Belle~II~\cite{Belle-II:2018jsg,Banerjee:2022vdd} \\
             $\tau\to \mu e\bar e$ & $1.8\times 10^{-8}$~Belle~\cite{Hayasaka:2010np}  & $2.9 \times 10^{-10}$~Belle~II~\cite{Belle-II:2018jsg,Banerjee:2022vdd} 
       \\ \hline
       $\tau\to e\gamma$ & $3.3\times 10^{-8}$~BaBar~\cite{BaBar:2009hkt} & $9.0\times 10^{-9}$~Belle~II~\cite{Belle-II:2018jsg,Banerjee:2022vdd} \\
              $\tau\to ee\bar e$ & $2.7\times 10^{-8}$~Belle~\cite{Hayasaka:2010np}  & $4.7 \times 10^{-10}$~Belle~II~\cite{Belle-II:2018jsg,Banerjee:2022vdd} \\
                    $\tau\to e\mu \bar \mu$ & $2.7\times 10^{-8}$~Belle~\cite{Hayasaka:2010np}  & $4.5 \times 10^{-10}$~Belle~II~\cite{Belle-II:2018jsg,Banerjee:2022vdd} \\
       \hline\hline
    \end{tabular}
    \caption{Present and expected future 90\%~CL limits on branching ratios of LFV decays. Notice that if $\ell = \ell^-$ then $\bar\ell = \ell^+$ and vice versa. For the modes involving invisible ALPs, we refer to discussions and references in Section~\ref{sec:sensitivity}.}
    \label{tab:decay_LFV}
\end{table}

The LFV decays giving the most stringent constraints on the paramter space that we are interested in include $\ell_i\to\ell_j\,a$, $\ell_i\to\ell_j\gamma$, $\ell_i\to\ell_j\ell_k\ell_k$.
The present and future experimental limits for these processes are collected in Table~\ref{tab:decay_LFV}.

If both LFV and LFC ALP interactions are present, they can induce the radiative LFV decays $\ell_i\to\ell_j\gamma$ at one loop. The resulting decay width reads~\cite{Cornella:2019uxs}:
\begin{align}
    \Gamma(\ell_i\to\ell_j\gamma)=\frac{\alpha \,m_{\ell_i}}{2}\bigg(|F_2(x_i)|^2+|G_2(x_i)|^2\bigg),
\end{align}
where $x_i \equiv m_a^2 / m_{\ell_i}^2$ and
\begin{align}
    F_2(x_i)=-\frac{m_{\ell_i} C_{\ell_i\ell_j}^A}{64\pi^2f_a^2}\bigg[m_{\ell_i}C_{\ell_i\ell_i}^Ag_1(x_i)+m_{\ell_j}C_{\ell_j\ell_j}^Ag_2(x_i)\bigg],\\
    G_2(x_i)=-\frac{m_{\ell_i} C_{\ell_i\ell_j}^V}{64\pi^2f_a^2}\bigg[m_{\ell_i}C_{\ell_i\ell_i}^Ag_1(x_i)-m_{\ell_j}C_{\ell_j\ell_j}^Ag_2(x_i)\bigg].
\end{align}
%
The loop functions $g_{1,2}$ are listed in Appendix~\ref{app:loop}.

If kinematically allowed, LFV ALP couplings alone induce lepton decays into ALPs, whose decay width reads in the limit $m_{\ell_j}\ll m_{\ell_i}$~\cite{Calibbi:2020jvd,Cornella:2019uxs}:
\begin{align}
    \Gamma(\ell_i\to\ell_j a)= \frac{1}{64\pi}\frac{m_{\ell_i}^3}{f_a^2}\left(|C_{\ell_i\ell_j}^V|^2+|C_{\ell_i\ell_j}^A|^2\right) \left(1-\frac{m_a^2}{m_{\ell_i}^2} \right)^2.
\end{align}

If LFC couplings exist, the 3-body decays $\ell_i\to\ell_j\ell_k\ell_k$ are also induced. When $2\,m_{\ell_k}<m_a<m_{\ell_i}-m_{\ell_j}$, on-shell ALP decays give the dominant contribution to these processes and one gets in the narrow width approximation:
\begin{align}
\Gamma(\ell_i\to\ell_j\ell_k\ell_k)=\Gamma(\ell_i\to\ell_j a)\,\text{BR}(a\to \ell_k\ell_k),
\end{align}
where the width of the process $a\to \ell_k\ell_k$ is given in
Eq.~\eqref{eq:atoellell}. 
In order to estimate the fraction of ALPs decaying inside the detector, thus giving rise to the 3-body signal, we follow Ref.~\cite{Cornella:2019uxs} and multiply the above expression by a factor $\bigg(1-\exp(-\frac{L}{l_a})\bigg)$, where $L\sim 1$~m is the typical radius of the detector in LFV experiments and $l_a$ is the mean flight length of the ALP, which is given by
\begin{align}
    l_a=\frac{p_a}{m_a \Gamma_a}\hbar c \,,\quad\quad 
    p_a=\frac{\lambda^{\frac{1}{2}}(m_{\ell_i}^2,m_{\ell_j}^2,m_a^2)}{2m_{\ell_i}},
\end{align}
with $\lambda(x,y,z) = x^2+y^2+z^2-2(xy+yz+zx)$.

If $m_a>m_{\ell_i}-m_{\ell_j}$, the 3-body decays can be still mediated by an off-shell ALP mediated. Neglecting the loop-suppressed ALP-photon coupling induced by the LFC interactions, one obtains the following expression~\cite{Cornella:2019uxs}:
\begin{align}
    \Gamma(\ell_i\to \ell_j a^*\to \ell_j\ell_k\ell_k)=\frac{|C_{\ell_k\ell_k}^A|^2 \left(|C_{\ell_i\ell_k}^V|^2+|C_{\ell_i\ell_k}^A|^2\right)}{512\pi^3}\frac{m_{\ell_i}^3m_{\ell_k}^2}{f_a^4}\phi_0^{jk}(x_i),
    \label{eq:muegamma}
\end{align}
where $x_i \equiv m_a^2 / m_{\ell_i}^2$ and
\begin{align}
    \phi^{jj}_0(x)=&-\frac{11}{4}+4 x-\left(\frac{1}{2} x^2 \log \left(\frac{2 x-1}{x}\right)-1+5 x-4 x^2\right) \log(\frac{x-1}{x})\nonumber\\
    &+\frac{1}{2} x^2 \left(\text{Li}_2 \left(\frac{x-1}{2 x-1}\right)-\text{Li}_2\left(\frac{x}{2 x-1}\right)\right)\,\\
    \phi^{j\ne k}_0(x)&=\left(3 x^2-4 x+1\right)\log \left(\frac{x-1}{x}\right)+3 x-\frac{5}{2}\,.
\end{align}

In addition, the same loops inducing $\ell_i \to \ell_j \gamma$ also contribute to $\ell_i\to \ell_j\ell_k\ell_k$ through a virtual photon exchange~\cite{Cornella:2019uxs,Kuno:1999jp}:
\begin{align}
     \Gamma(\ell_i\to \ell_j \gamma^*\to \ell_j\ell_k\ell_k) \simeq \frac{\alpha}{3\pi} \left[\log\left(\frac{m_{\ell_i}}{m_{\ell_j}}\right)^2-3+\frac{\delta_{jk}}{4}\right]\, \Gamma(\ell_i\to \ell_j \gamma)\,.
\end{align}

\subsection{Anomalous magnetic moments}
\label{app:g-2}
The ALP-induced contribution $\Delta a_\ell$ to the anomalous magnetic dipole moments of the leptons\,---\,defined as $a_\ell = (g_\ell -2)/2$\,---\,can have two parts, depending on whether the LFC couplings are present or not~\cite{Bauer:2021mvw,Cornella:2019uxs}. The contribution from the LFC couplings is given by
\begin{align}
   \Delta a_{\ell_i}^{\text{LFC}}=-\frac{m_{\ell_i}^2}{16\pi^2f_a^2}\left[ (C_{\ell_i\ell_i}^A)^2h_1(x_i)+16\pi\alpha \,C_{\ell_i\ell_i}^AC_{\gamma\gamma}^{\text{loop}}\left(\log(\frac{4f_a^2}{m_{\ell_i}^2})-h_2(x_i)\right)\right],
   \label{eq:a_LFC}
\end{align}
where the loop functions $h_{1,2}$ are given in Appendix~\ref{app:loop} and $x_i={m_a^2}/{m_{\ell_i}^2}$. The LFC couplings also induce the $a\gamma\gamma$ coupling at one loop\,---\,see Appendix~\ref{app:gamma}\,---\,such that $C_{\gamma\gamma}^{\text{loop}}$ reads
\begin{align}
  &C_{\gamma\gamma}^{\text{loop}}=\frac{1}{8\pi^2}\sum_i C_{\ell_i\ell_i}^A B\left(\frac{4}{x_i}-i\epsilon\right)\,,
\end{align}
where the loop function $B$ is given in Eq.~\eqref{eq:B}.

The LFV contribution is given by
\begin{align}
    \Delta a_{\ell_i}^{\text{LFV}}= \frac{m_{\ell_i} m_{\ell_k}}{32\pi^2f_a^2}\left\{\left[|C_{\ell_i\ell_k}^V|^2+|C_{\ell_i\ell_k}^A|^2\right]\left[\frac{m_{\ell_i}^3}{m_{\ell_k}^3}\int_0^1\dd x\frac{x(1-x)^2}{\Delta_{i\to i}}+\frac{m_{\ell_i}}{m_{\ell_k}}\int_0^1\dd x\frac{(1-x)^2(x-2)}{\Delta_{i\to i}}\right]\right.\nonumber\\
    \left.\left[|C_{\ell_i\ell_k}^V|^2-|C_{\ell_i\ell_k}^A|^2\right]\left[\int_0^1\dd x\frac{(1-x)^2}{\Delta_{i\to i}}+\frac{m_{\ell_i}^2}{m_{\ell_k}^2}\int_0^1\dd x\frac{(1-x)^2(1-2x)}{\Delta_{i\to i}}\right]\right\}\,,
    \label{eq:a_LFV}
\end{align}
where $\Delta_{i\to i}=x x_k+x(x-1){m_{\ell_i}^2}/{m_{\ell_k}^2}+1-x$. In this paper, we focus on $V\pm A$ interactions, hence the second line vanishes. Then, in the limit $m_{\ell_i}\ll m_{\ell_k}$,
\begin{align}
    \Delta a_{\ell_i}^{\text{LFV}}=\frac{m_{\ell_i}^2[|C_{\ell_i\ell_k}^V|^2+|C_{\ell_i\ell_k}^A|^2]}{32\pi^2f_a^2}\frac{6 (1-2 x_k) x_k^2 \log(x_k)+(x_k-1) ((20 x_k-19) x_k+5)}{6 (x_k-1)^4}\,,
\end{align}
while in the limit $m_{\ell_i}\gg m_{\ell_k}$
\begin{align}
\Delta a_{\ell_i}^{\text{LFV}}=\frac{m_{\ell_i}^2[|C_{\ell_i\ell_k}^V|^2+|C_{\ell_i\ell_k}^A|^2]}{32\pi^2f_a^2} \left( x_i^2\log(\frac{x_i}{x_i-1})-x_i-\frac{1}{2}\right)\,.
\end{align}

\begin{figure}[t]
    \centering
    \subfigure[LFV ($e$-$\mu$) couplings only]{
    \includegraphics[width=0.49\textwidth]{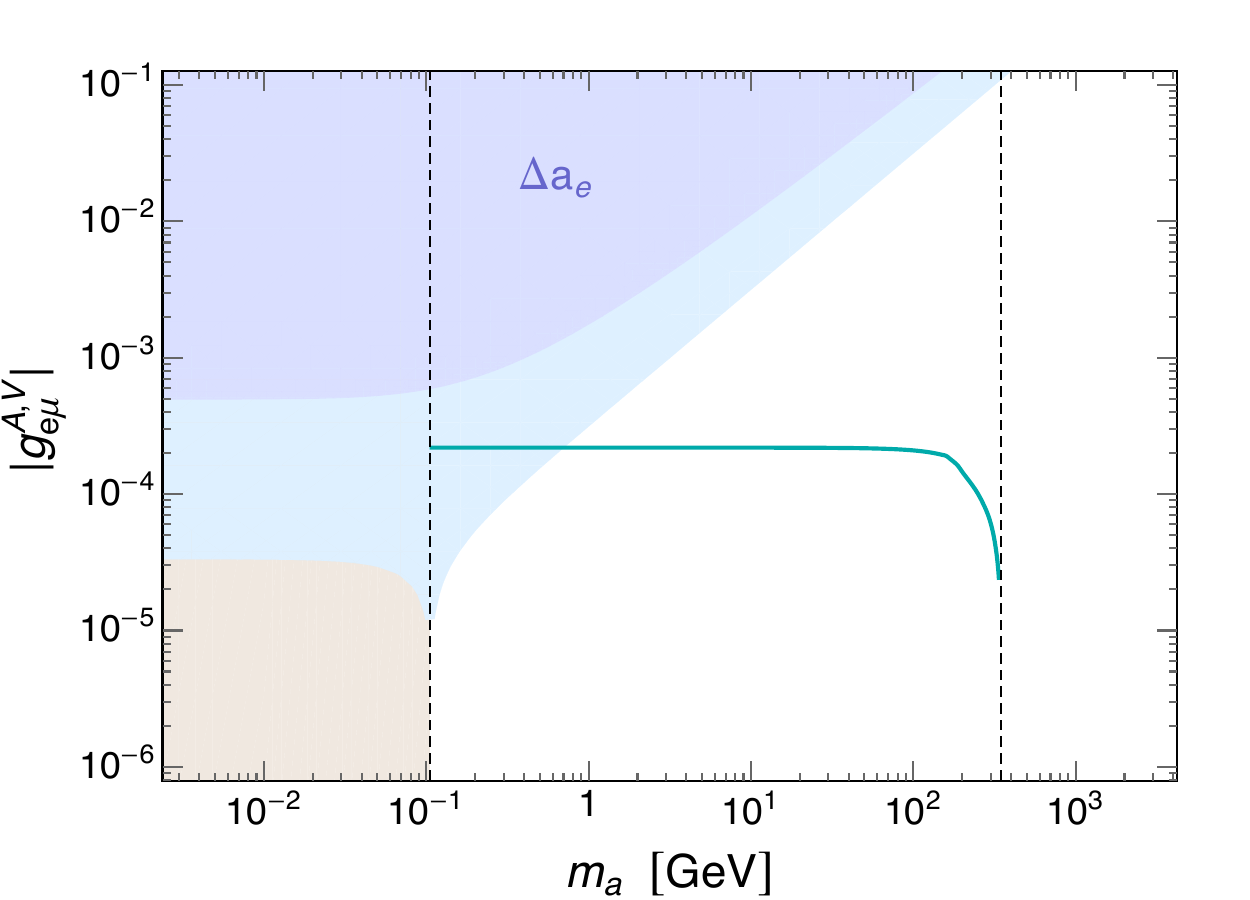}}
    \hfill
    \subfigure[LFV ($e$-$\mu$) and LFC couplings]{
    \includegraphics[width=0.49\textwidth]{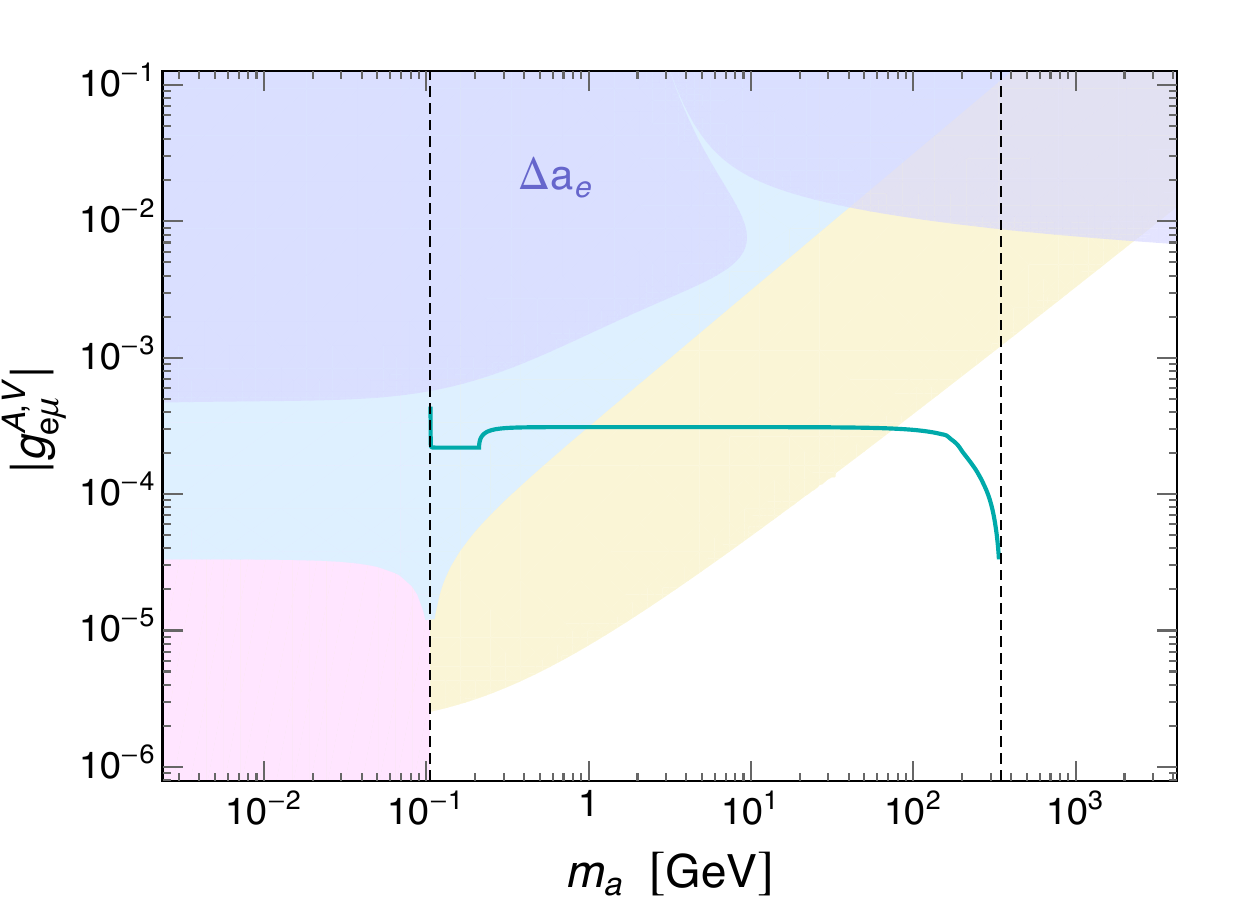}}
    \hfill
    \subfigure[LFV ($e$-$\mu$) couplings only]{
    \includegraphics[width=0.49\textwidth]{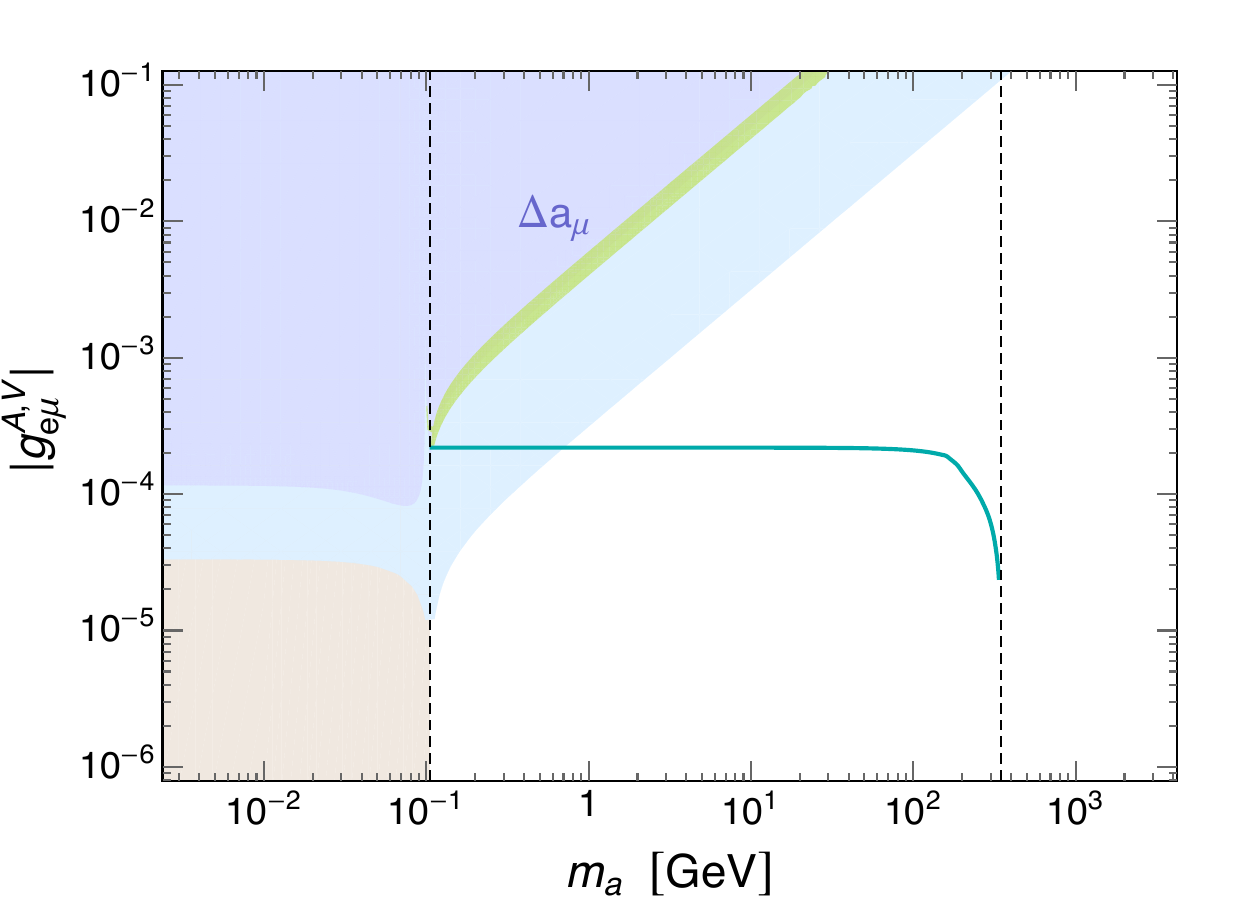}}
    \hfill
    \subfigure[LFV ($e$-$\mu$) and LFC couplings]{
    \includegraphics[width=0.49\textwidth]{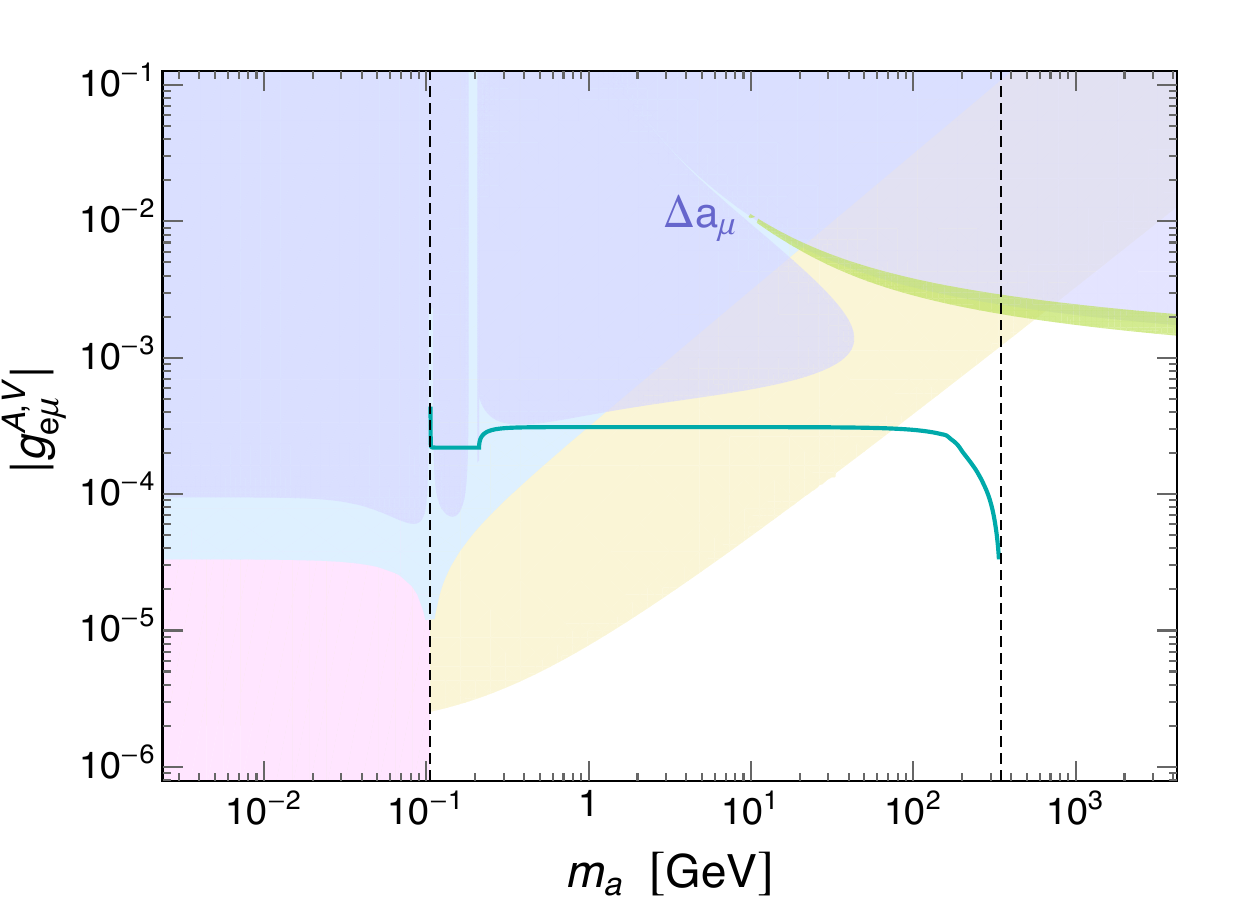}}
    \caption{Electron (first row) and muon (second row) $g-2$ constraints on the same parameter space regions depicted in Figure~\ref{fig:constraint_gemu_zoom_to_collider}. Purple areas correspond to ALP contributions $\Delta a_\ell$ beyond the conservative $2\sigma$ ranges discussed in the text. In the green regions, the ALP can explain the discrepancy between the measured  value of $a_\mu$ and the SM prediction of the WP~\cite{Aoyama:2020ynm} at the $2\sigma$ level.}
    \label{fig:g-2}
\end{figure}

As mentioned in Section~\ref{sec:sensitivity}, the SM predictions of both electron and muon $g-2$ are currently under scrutiny. In order for $a_e$ to test NP effects, theoretical calculations of the SM contributions must be supplemented by a very precise measurement of the fine-structure constant $\alpha$ (that is, with an accuracy of $\approx 0.1$ parts per billion)~\cite{Giudice:2012ms}. In recent years, this has become possible through atomic matter-wave interferometry. However, the two most precise measurements to date, which respectively employed  cesium~(Cs)~\cite{Parker:2018vye} and rubidium~(Rb)~\cite{Morel:2020dww} atoms, are in strong mutual tension. Depending on which result is adopted, the comparison between SM prediction and the experimental measurements~\cite{Fan:2022eto,Hanneke:2008tm} yields
\begin{align}
    \Delta a_e^{\text{Cs}} &~\equiv a_e^\text{Cs} - a_e^\text{exp} = -(8.8\pm3.6)\times10^{-13}\,, \\
    \Delta a_e^{\text{Rb}} &~\equiv a_e^\text{Rb} - a_e^\text{exp} = (4.8\pm3.0)\times10^{-13}\,.
    \label{eq:Rb}
\end{align}
As we can see, the Cs-based prediction is in mild tension with the experimental measurement, while the Rb one is in good agreement. Therefore, we impose throughout the paper that the ALP contributions to $a_e$ do not exceed the $2\sigma$ range of Eq.~\eqref{eq:Rb}.

As is well known, the SM prediction of $a_\mu$ is affected by a certain tension between the lattice QCD calculation of the leading hadronic contribution performed by the BMW collaboration~\cite{Borsanyi:2020mff} and that based on dispersion relations and data of hadron production in low-energy $e^+e^-$ collisions, as summarised by the 2020 white paper~(WP)~\cite{Aoyama:2020ynm}. The two calculations have comparable precision and, once compared with the latest measurement of the Muon g-2 collaboration~\cite{Muong-2:2023cdq}, give
\begin{align}
    \label{eq:wp} 
    \Delta a_\mu^{\text{WP}} &~\equiv a_\mu^\text{WP} - a_\mu^\text{exp} = (2.49\pm 0.48)\times10^{-9}\,, \\
    \Delta a_\mu^{\text{BMW}} &~\equiv a_\mu^\text{BMW} - a_\mu^\text{exp} = (1.05\pm 0.61)\times10^{-9}\,.
    \label{eq:bmw}
\end{align}
The first range would exclude most of the parameter space of our ALP models, especially the regions predicting negative values of $\Delta a_\mu$. However, it does exclude the SM itself at the $5\sigma$ level. Since it is not possible at this stage to claim that a NP contribution of this size is indeed necessary, we again adopt a conservative approach and use Eq.~\eqref{eq:bmw} as a constraint in our analysis. As one can see, this can exclude both positive and negative ALP contributions $\Delta a_\mu$ if too large in absolute value.

The impact of the $g-2$ constraints on the regions of the parameter space studied in Section~\ref{sec:mueLFV} is shown in Figure~\ref{fig:g-2}. The purple regions are outside the $2\sigma$ ranges of Eqs.~\eqref{eq:Rb}
and~\eqref{eq:bmw}, while the green stripes correspond to values of $\Delta a_\mu$ \emph{within} the $2\sigma$ range preferred by Eq.~\eqref{eq:wp}. 
We can see that, if the ALP only couples to $e$-$\mu$ (left plots), neither $\Delta a_e$ nor $\Delta a_\mu$ exclude portions of our parameter space beyond the limits from searches for $\text{Mu}-\overline{\text{Mu}}$ oscillations (blue region).
Notice that here we plot the rather special case $C^A_{e\mu} = C^V_{e\mu}$, that is, a model with ALPs only coupling with right-handed leptons, for which the second line of Eq.~\eqref{eq:a_LFV} vanishes. However, we checked that the above conclusion is barely affected even if left-handed couplings are also present (e.g.~setting $C^V_{e\mu}=0$, $C^A_{e\mu}\neq 0$), consistently with the results of Ref.~\cite{Endo:2020mev}. 

If the ALP also interacts with LFC currents (right plots), we see that the $g-2$ constraints become stronger and there is the possibility of changes of sign and multiple cancellations between the contribution in Eq.~\eqref{eq:a_LFV} and the two terms of Eq.~\eqref{eq:a_LFC}. Still, this does not imply additional constraints beyond the limit from $\mu\to e\gamma$ (yellow area) on the region that can be tested by $\mu$TRISTAN. As we can see, this is not the case for ALPs that are too heavy to be produced at $\mu$TRISTAN, for which the leptonic $g-2$, in particular $\Delta a_\mu$, can provide the most important bounds on the ALP couplings.

\subsection{Loop functions}
\label{app:loop}

The expressions for the loop functions appearing in this section are the following~\cite{Cornella:2019uxs}:
\begin{align}
    &g_1(x)=\frac{(x-3) x^2 \log x}{x-1}-2 x+1-2 \sqrt{x-4} x^{\frac{3}{2}} \log \left(\frac{\sqrt{x-4}+\sqrt{x}}{2} \right),\\
    &g_2(x)=1-2 x+2 (x-1) x \log \left(\frac{x}{x-1}\right),\\
    &h_1(x)=1+2 x-(x-1) x \log x+2 (x-3) \sqrt{\frac{x}{x-4}} x \log \left(\frac{\sqrt{x-4}+\sqrt{x}}{2} \right),\\
    &h_2(x)=1+\frac{1}{6} x^2 \log x-\frac{x}{3}-\frac{1}{3} (x+2) \sqrt{x (x-4)} \log \left(\frac{\sqrt{x-4}+\sqrt{x}}{2} \right).
\end{align}


\bibliographystyle{JHEP}
\bibliography{ref}

\end{document}